\documentclass[a4paper,11pt]{article}

\usepackage{bm}
\usepackage{anysize}
\usepackage{amsthm}
\usepackage{amssymb}
\usepackage{amsmath}%
\usepackage{fancybox}
\usepackage{bbm}
\usepackage{amsfonts}
\usepackage[T1]{fontenc}
\usepackage{latexsym}
\usepackage{makeidx}
\usepackage{cite}  
\usepackage[numbers]{natbib}
\usepackage{mathabx}

\usepackage{tikz}
\usepackage{float}
\usepackage{subcaption}

\usepackage[all]{xy}

\usepackage{stackengine}
\stackMath

\newtheorem{teo}{Theorem}
\newtheorem{lem}{Lemma}
\newtheorem{pro}{Proposition}
\newtheorem{cor}{Corollary}
\newtheorem{defi}{Definition} 
\newtheorem{remark}{Remark}[section]

\usepackage{mathtools}

\DeclarePairedDelimiter\floor{\lfloor}{\rfloor}

\newcommand*{\m}[1]{\underline{#1}}
\newcommand{\un}[1]{\underline{#1}}

\newcommand{\fd}{\rightarrow}

\newcommand{\inc}{\subset}

\newcommand{\iso}{\cong}

\newcommand{\ba}{\overline}

\newcommand{\al}{\alpha}
\newcommand{\be}{\beta}

\newcommand{\fhi}{\varphi}
\newcommand{\del}{\delta}
\newcommand{\Del}{\Delta}

\newcommand{\Gam}{\Gamma}

\newcommand{\Om}{\Omega}

\newcommand{\N}{\mathbb{N}}

\newcommand{\R}{\mathbb{R}}
\newcommand{\C}{\mathbb{C}}

\newcommand{\E}{\mathbb{E}}

\newcommand{\I}{\mathbbm{1}}

\newcommand{\Sa}{\mathbb{S}}

\newcommand{\pa}{\partial}

\newcommand{\el}{\ell}

\def\pf{\par\noindent {\em Proof.}~\par\noindent}

\def\lim{\mathop{\mbox{\normalfont lim}}\limits}

\def\pf{\par\noindent {\em Proof. }}

\def\pa{\partial}

\begin{document}

\date{}

\title{Pizzetti formulae and the Radon Transform on the Sphere}
\small{
\author
{Al\'i  Guzm\'an Ad\'an$^\dagger$\thanks{Postdoctoral Fellow of the Research Foundation - Flanders (FWO)} \and Mihaela B. Vajiac$^\ddagger$}
\vskip 1truecm
\date{\small  $^\dagger$Clifford Research Group, Department of Mathematical Analysis, Faculty of Engineering
and Architecture, Ghent University, Krijgslaan 281, 9000 Gent, Belgium. \\
{\tt Ali.GuzmanAdan@UGent.be},\\[+.2cm]
$^\ddagger$Schmid College of Science and Technology\\
Chapman University, One University Drive, Orange, CA 92866\\
{\tt mbvajiac@chapman.edu} }

\maketitle

\begin{abstract} 
In this paper, we obtain Pizzetti-type formulae on regions of the the unit sphere $\mathbb{S}^{m-1}$ of $\mathbb{R}^m$, and study their applications to the problem of inverting the spherical Radon transform. In particular, we approach integration over $(m-2)$-dimensional sub-spheres of $\mathbb{S}^{m-1}$, $(m-1)$-dimensional sub-balls, and over $(m-1)$-dimensional spherical caps as the action of suitable concentrated delta distributions. In turn, this leads to Pizzetti formulae that express such integrals in terms of the action of SO$(m-1)$-invariant differential operators. In the last section of the paper, we use some of these expressions to derive the inversion formulae for the Radon transform on $\mathbb{S}^{m-1}$ in a direct way.


\noindent

\vspace{0.3cm}

\small{ }

\noindent
\textbf{}
\end{abstract}


\section{Introduction}

The classical Pizzetti formula expresses the integral over the unit sphere $\Sa^{m-1}\inc\R^m$
as a certain power series of the Euclidean Laplacian operator acting on
the integrand \cite{Pizz}. This fits in the larger framework of Stiefel manifolds recently studied by
Coulembier and Kieburg in \cite{MR3395222}, as $\Sa^{m-1}\iso\textup{SO}(m)/\textup{SO}(m-1)$  is the Stiefel manifold of order $k=1$ in $\R^m$. 

In \cite{MR3395222, Ali_Pizz}, Pizzetti formulae for the Stiefel manifold 
\[
\textup{St}(m, k):=\left\{(\m{\omega}_1\, \ldots, \m{\omega}_k)\in \left(\R^m\right)^k: \langle\m{\omega}_j, \m{\omega}_\el\rangle=\del_{j,\el}\right\} \iso \textup{SO}(m)/\textup{SO}(m-k)
\]
were derived in terms of a recursive application of {$k$} Pizzetti formulae over geodesic sub-spheres of $\Sa^{m-1}$ of respective codimensions $1, 2, \ldots, k$ ($1\leq k\leq m-2$). These formulae turn out the be closely related to the Radon transform on $\Sa^{m-1}$. This relation becomes evident from the fact that spherical Radon transforms can be considered as functions on Stiefel manifolds, and reciprocally, integrals on Stiefel manifolds can be written in terms of Radon transforms. 

Indeed, we recall that the spherical Radon transform maps integrable functions on $\Sa^{m-1}$ to functions defined on the set $\Xi$ of $(m-k-1)$-dimensional totally geodesic sub-spheres of $\Sa^{m-1}$. This transformation takes place by means of the correspondence 
\[
f \mapsto \hat{f},  \;\;\;\;\; \mbox{ with} \;\;\;\;\; \hat{f} (\xi) = \int_{\xi} f(\underline{x})d S_{\underline{x}},
\]
where $f\in L^2(\Sa^{m-1})$, $\xi\in  \Xi$, and $d S$ is the $(m-k-1)$-dimensional Euclidean surface measure on $\xi$. It is easily seen that $\hat{f}$ can be written as a function on St$(m,k)$ as follows 
\begin{align}\label{RadvsStief}
\hat{f}(\m{\omega}_1, \ldots, \m{\omega}_k) &= 2\int_{\R^m} \del\left(\|\m{x}\|^2-1\right) \del(\langle\m{x},\m{\omega}_1\rangle) \ldots \del(\langle\m{x}\, ,\m{\omega}_k\rangle) f(\m{x})\, dV_{\m{x}}\, ,  &(\m{\omega}_1, \ldots, \m{\omega}_k) &\in \textup{St}(m,k)\, ,
\end{align}
{where $dV_{\m{x}} = dx_1\cdots dx_m$ is the classical Lebesgue measure in $\R^m$. Here} we have made use of the concentrated delta distribution  $\del\left(\|\m{x}\|^2-1\right) \del(\langle\m{x},\m{\m{\omega}_1}\rangle) \ldots \del(\langle\m{x},\m{\m{\omega}_k}\rangle)$ (see Section \ref{CDDSec}), and of the mapping $\fhi: \textup{St}(m,k)\fd \Xi$ given by
\begin{equation}\label{St-SubS}
(\m{\omega}_1, \ldots, \m{\omega}_k) \mapsto \xi=\left\{\m{x}\in\Sa^{m-1}: \langle\m{x},\m{\omega}_1\rangle=\cdots = \langle\m{x},\m{\omega}_k\rangle=0\right\}.
\end{equation}
In (\ref{RadvsStief}) we have abused of the notation $\hat{f}$ to refer to the function $\hat{f}\circ \fhi$.  As a matter of fact,  following the same procedure, (\ref{St-SubS}) allows one to write any function on $\Xi$ as a function defined on $\textup{St}(m,k)$.

{Similarly, it is possible to connect integration over St$(m,k)$ with $(m-k)$-dimensional spherical Radon transforms. Indeed, given a function $\Phi(\m{\omega}_1, \ldots, \m{\omega}_k)$ of $k$ vector variables in $\R^m$, one easily obtains 
\begin{equation}\label{StiefvsRad}
\int_{\textup{St}(m,k)} \Phi(\m{\omega}_1, \ldots, \m{\omega}_k) \, dS_{\m{\omega}_1\, \ldots, \m{\omega}_k} = \int_{\textup{St}(m,k-1)} \hat{\Phi}(\m{\omega}_1\, \ldots, \m{\omega}_{k-1}) \, dS_{\m{\omega}_1, \ldots, \m{\omega}_{k-1}}, 
\end{equation}
where $ \hat{\Phi}(\m{\omega}_1\, \ldots, \m{\omega}_{k-1})$ denotes the $(m-k)$-dimensional Radon transform of the function $\Phi(\m{\omega}_1, \ldots, \m{\omega}_{k-1}, \cdot)$ evaluated at the subsphere orthogonal to the $(k-1)$ orthonormal frame  $(\m{\omega}_1\, \ldots, \m{\omega}_{k-1})$, i.e.\
\[
\hat{\Phi}(\m{\omega}_1\, \ldots, \m{\omega}_{k-1}) = 2 \int_{\R^m}  \del\left(\|\m{w}_k\|^2-1\right) \del(\langle\m{w}_1,\m{\omega}_k\rangle) \ldots \del(\langle\m{w}_k\, ,\m{\omega}_k\rangle) \Phi(\m{\omega}_1, \ldots, \m{\omega}_k)\, dV_{\m{w}_k} \,.
\]
}


This paper is intended to be the first of a series of manuscripts dedicated to study the applications of Pizzetti-type formulae to the problem of inverting spherical Radon transforms, by exploiting the relations (\ref{RadvsStief}) and (\ref{StiefvsRad}). An additional motivation for this idea is the fact that the dual of the Radon transform, which plays a central role in the inversion formulae, can also be written as an integral on a Stiefel manifold. Indeed, the dual Radon transform of a continuous function $\phi$ on $\Xi$ is defined on unit vectors $\m{x}\in\Sa^{m-1}$ as as the average of $\phi$ over the set of geodesic sub-spheres $\xi\in\Xi$ passing through $\m{x}$, i.e.
\[
\widecheck{\phi}(\m{x}) = \int_{\xi\ni\m{x}} \phi(\xi) \, d\mu(\xi),
\]
where $d\mu$ is the normalized invariant measure on the set $\{\xi\in\Xi: \m{x}\in\xi\}$. We recall that this set is invariant under the action of the group of rotations around $\m{x}$. In view of the mapping (\ref{St-SubS}), the above integral can be rewritten as the following integral over $\textup{St}(m-1,k)$, {seeing the latter} as the set of all orthonormal $k$-frames orthogonal to $\m{x}$, i.e.\
\begin{equation}\label{dualStf}
\widecheck{\phi}(\m{x}) = \frac{1}{\textup{vol}(\textup{St}(m-1,k))} \int_{\textup{St}(m,k)} \del(\langle\m{x},\m{\omega}_1\rangle) \ldots \del(\langle\m{x},\m{\omega}_k\rangle) \, \phi(\m{\omega}_1, \ldots, \m{\omega}_k) \, dS_{\m{\omega}_1\, \ldots, \m{\omega}_k}.
\end{equation}
In particular, it is our goal to find alternative direct proofs for the following inversion theorems for the Radon transform on $\Sa^{m-1}$, which have been proved in \cite{MR754767, MR2743116}. 
\begin{teo}\label{Helgason spherical inversion, general k even}
\textup{\cite[Thm.~1.17 Ch.~3.1]{MR2743116}.}
Let $m,k\in \N$ be natural numbers such that $1\leq k\leq m-2$. Assume $m-k-1$ even and let $P_{m-k-1}$ be the polynomial
\[
P_{m-k-1}(z)=\prod_{j=1}^{\frac{m-k-1}{2}} \left[z-(m-k-2j)(k+2j-2)\right]\, ,
\]
of degree $\frac{m-k-1}{2}$. The $(m-k-1)$-dimensional Radon transform on $\Sa^{m-1}$ $f\longrightarrow \widehat{f}$ is, for even functions $f$, inverted by the formula
\begin{align}\label{k even genral}
2(-4\pi)^{\frac{m-k-1}{2}} \frac{\Gam\left(\frac{m-1}{2}\right)}{\Gam\left(\frac{k}{2}\right)} \, f=P_{m-k-1}(\Delta_{LB}) \; (\widehat{f} \,) \widecheck{\phantom{f}},
\end{align}
where $\Delta_{LB}$ is the Laplace-Beltrami operator on $\mathbb S^{m-1}$ (see Section \ref{pre}).
\end{teo}

\begin{teo}\label{Helgason2general}
\textup{\cite[Thm.~1.22 Ch.~3.1]{MR2743116}.}
The $(m-k-1)$-dimensional Radon transform $f\mapsto \widehat{f}$ on $\Sa^{m-1}$ is, for even functions $f$,   inverted by
\begin{align}\label{k general}
f(\underline{x})=\frac{2^{m-k-1}}{(m-k-1)!\, \sigma_{m-k}} \, \left(\frac{d}{dt^2}\right)^{m-k-1} \left[ \int_0^t {(\widehat{f}\,)}^{\vee}_{\cos^{-1}(q)}(\underline{x}) \, q^{m-k-1} \, (t^2-q^2)^{\frac{m-k-3}{2}}\, dq \right]_{t=1},
\end{align}
where by $\widecheck{\phi}_r(\underline{x})$ we denote the generalization of the dual Radon transform defined in (\ref{avg}) in Section \ref{SubInvGenk1}.
\end{teo}
In this paper, we will restrict our attention to the case of codimension $k=1$. We will use this case as a starting point in our study because it is the simplest-case scenario to consider from a computational point of view. Indeed, formula (\ref{RadvsStief}) shows that the $(m-2)$-dimensional Radon transform can be seen as an even function defined on $\Sa^{m-1}$.
Moreover, formula (\ref{dualStf}) shows that the dual of this transform is also an integral over $(m-2)$-dimensional geodesic sub-spheres, which makes it (up to a constant) identical to the Radon transform. Hence, in order to deal with the inversion formulae given in Theorems \ref{Helgason spherical inversion, general k even} and \ref{Helgason2general} for $k=1$, we only need to establish Pizzetti-type ones for all (not necessary geodesic) subspheres on $\Sa^{m-1}$, which is achieved in Theorem \ref{PizzMainTheo}. In a forthcoming paper, we shall study the more general case of arbitrary codimension $1\leq k\leq m-2$. 

The advantage of using Pizzetti formulae to prove these inversion results is that they offer a direct, straightforward way of computing the expressions on the right hand sides of (\ref{k even genral}) and (\ref{k general}). However, it is worth mentioning that this computational method hides, up to some extend, the geometric properties used by Helgason in his original proofs in \cite{MR754767, MR2743116}. This is due to the fact that the Pizzetti formulae implicitly  encode the group invariance of the sub-spheres under the action of suitable rotation subgroups of SO$(m)$. 

The paper is structured as follows. In Section~\ref{pre} we briefly present the notation, definitions and preliminary results needed in the sequel. We pay particular attention to the notions spherical harmonics in $\R^m$, Clifford algebras, and the use of concentrated delta distributions in integration. Section  \ref{Pizzetti on regions of sphere} is devoted to the study of Pizzetti-type formulae describing integration on certain regions of the unit sphere $\Sa^{m-1}$. In particular, we consider the integral over $(m-2)$-dimensional sub-spheres, $m$-dimensional sub-balls, and $(m-1)$-dimensional spherical caps. These results are all summarized in Theorem \ref{PizzMainTheo}. The integral results found for the spherical caps yield a new expression for the  classical Pizzetti formula on $\Sa^{m-1}$ which is discussed in more detail in Appendix \ref{Appendix}. Finally, in Section ~\ref{Radon Sperical}, we apply the Pizzetti formulae found for the $(m-2)$-dimensional sub-spheres to prove Helgason's  Radon inversion results 
 given in Theorems  \ref{Helgason spherical inversion, general k even} and \ref{Helgason2general} for $k=1$.



\section{Preliminaries}
\label{pre}
In this section we provide a few preliminaries on the topics of spherical harmonics, Clifford algebras, and integration over manifolds using distributions; which shall be useful in the sequel. 
\subsection{Spherical harmonics}

Let $\mathcal{P}(\R^m)=\C[x_1, \ldots, x_m]$ be the space of complex-valued polynomials in the vector variable $\m{x}=(x_1, \ldots, x_m)^T$ of $\R^m$. On $\R^m$ we consider the standard Euclidean inner-product $\langle\m{x}, \m{y}\rangle= \sum_{j=1}^mx_jy_j$ and its associated norm $\|\m{x}\|=\sqrt{\langle\m{x}, \m{x}\rangle}$. 

A polynomial $P \in \mathcal{P}(\R^m)$ is called {\em homogeneous of degree $k \in \N_0:=\N\cup\{0\}$ (or  $k$-homogeneous)} if it holds for every $\m{x}\in \R^m\setminus\{0\}$ that
\[
P(\m{x}) = \|\m{x}\|^k P\left(\frac{\m{x}}{\|\m{x}\|}\right), 
\]
where $\frac{\m{x}}{\|\m{x}\|}$ is the restriction of $\m{x}$ to the unit sphere $\Sa^{m-1}:=\{\m{\omega}\in\R^m: \|\m{\omega}\|=1\}$. 

It is well-known that $k$-homogeneous polynomials are the eigenfunctions of the Euler operator $\E=\sum_{j=0}^m x_j \pa_{x_j}$  corresponding to the eigenvalue $k$. 
Therefore one can define the space of homogeneous polynomials of degree $k$ as
\[
\mathcal{P}_k(\R^m) = \{P\in\mathcal{P}(\R^m): \E[P] = k P\},
\]
which allows for the decomposition
\begin{equation}\label{Dec1}
\mathcal{P}(\R^m) = \bigoplus_{k=0}^\infty \mathcal{P}_k(\R^m).
\end{equation}
\begin{defi}
The space of $k$-homogeneous polynomials that are also {\em harmonic}, (i.e.\ null-solutions of the Laplace
operator $\Del_{\m{x}} = \sum_{j=1}^{m} \pa^2_{x_j}$),  is denoted by 
\[
\mathcal{H}_k = \{P\in \mathcal{P}_k(\R^m): \Del_{\m{x}}[P]=0\}.
\]
The restriction of such $k$-homogeneous, harmonic polynomials to the unit sphere $\Sa^{m-1}$ are called {\em spherical harmonics of degree $k$}. We denote the space of all spherical harmonics of degree $k$ by $\mathcal{H}_k(\Sa^{m-1})$. 
\end{defi}

The spaces of spherical harmonics $\mathcal{H}_k(\Sa^{m-1})$ are eigenspaces of the Laplace-Beltrami operator on the sphere $\Sa^{m-1}$
\[
\Del_{LB} = \|\m{x}\|^2\Del_{\m{x}}-(m-2+\mathbb E)\mathbb E\, .
\]

 When acting on $\mathcal{P}(\R^m)$, the operators $\Del_{\m{x}}$, $\|\m{x}\|^2$ and $\E$ generate a representation of the special linear Lie algebra $\mathfrak{sl}_2$. Indeed, it can be easily verified that 
\begin{align*}
\left[\frac{\Del_{\m{x}}}{2}, \frac{\|\m{x}\|^2}{2}\right] &= \E + \frac{m}{2}, & \left[\frac{\Del_{\m{x}}}{2}, \E + \frac{m}{2}\right] &= {\Del_{\m{x}}}, & \left[\frac{\|\m{x}\|^2}{2}, \E + \frac{m}{2}\right] &= -\|\m{x}\|^2,
\end{align*}
where $[a,b]:=ab-ba$.

The {Laplacian} $\Del_{\m{x}}$ also plays an essential role when considering $\mathcal P(\R^m)$ as a representation of the special orthogonal group SO$(m)$ under the natural action $H:\textup{SO}(m)\fd \textup{Aut}(\mathcal P(\R^m))$ given by
\[
M[P](\m{x}) = P(M^{-1}\m{x}), \;\;\;\; M\in \textup{SO}(m), \; P\in \mathcal P(\R^m).
\]
Indeed, it is easily seen that $\Del_{\m{x}}$ commutes with the above action of $ \textup{SO}(m)$, which implies that $\mathcal{H}_k$ is a $\textup{SO}(m)$-invariant subspace of  $\mathcal P_k(\R^m)$. The main assertions of the classical theory of spherical harmonics can be summarized as follows (see, for example~\cite{MR754767, MR3060033}).
\begin{pro}\label{FisDec}{\bf [Fischer decomposition]}
\begin{itemize}
\item[i)] The spaces  $\mathcal{H}_k$  $(k\in\N_0)$ are irreducible representations of SO$(m)$.
\item[ii)] $\displaystyle \mathcal P_k(\R^m)= \mathcal{H}_k \oplus \|\m{x}\|^2 \mathcal P_{k-2}(\R^m)$.
\item[iii)] $\displaystyle  \mathcal P_k(\R^m)= \bigoplus_{j=0}^{\floor{\frac{k}{2}}}  \|\m{x}\|^{2j} \mathcal H_{k-2j}$.
\end{itemize}
\end{pro}
When considering the restrictions of polynomials to $\Sa^{m-1}$, the statement $iii)$ provides an orthogonal decomposition with respect to the inner product in $L^2(\Sa^{m-1})$
\[
\big\langle P(\m{x}), Q(\m{x})\big\rangle_\Sa = \frac{1}{\sigma_m} \int_{\Sa^{m-1}} {P(\m{x})^c} \, Q(\m{x}) \,dS_{\m{x}}\, ,
\]
where $dS_{\m{x}}$ is the {\em Lebesgue measure on} $\Sa^{m-1}$, $\sigma_m= \frac{2\pi^{\frac{m}{2}}}{\Gam\left(\frac{m}{2}\right)}$ is the  surface area of $\Sa^{m-1}$ and $(\cdot)^c$ stands for the complex conjugation. More in general, the decomposition in Proposition \ref{FisDec} $iii)$ can be extended to $L^2(\Sa^{m-1})$ as follows (see e.g. \cite[Ch.\ 1]{MR754767}).
\begin{pro}\label{OrthDecSH} Any function $f\in L^2(\Sa^{m-1})$ admits a unique decomposition into spherical harmonics,
\[
f(\m{\omega})= \sum_{k=0}^\infty H_{k}(\m{\omega})\, , \;\;\;\;\;\;\; H_{k} \in \mathcal H_{k}(\Sa^{m-1})\, .
\]
\end{pro}

\subsection{Clifford algebras}
{Throughout} this paper, we shall use the language of Clifford algebras as a tool to compute some of the geometric quantities needed in the sequel. Let $\{e_1,...,e_m\}$ be an orthonormal basis of $\R^m$. The {\em real Clifford algebra} $\mathbb{R}_m$ is the real associative algebra with generators $e_1,...,e_m$ satisfying the defining relations $e_je_{\el}+e_{\el}e_j=-2\delta_{j\el}$, for $j,\el=1,...,m$, where $\delta_{j\el}$ is the Kronecker symbol. Every element $a\in \mathbb{R}_m$ can be written in the form 
\[a=\sum_{A\inc M}a_Ae_A\, ,\]
where $a_A\in \R$, $M:=\{1,\ldots, m\}$ and for any multi-index $A=\{j_1, \ldots, j_k\}\subseteq M$ with $j_1< \ldots < j_k$ we put $e_A=e_{j_1}\cdots e_{j_k}$ and $|A|=k$. Every $a\in\mathbb{R}_m$ admits a multi-vector decomposition 
\[a=\sum_{k=0}^m [a]_k\, , \hspace{.5cm} \mbox{ where } \hspace{.5cm} [a]_k=\sum_{|A|=k}a_A e_A\, .\]
Here $[\cdot]_k:\R_m\fd \R_m^{(k)}$ denotes the canonical projection of $\R_m$ onto the space of $k$-vectors $\R_m^{(k)}=\textup{span}_\R\{e_A:|A|=k\}$ . Note that $\R_m^{(0)}=\R$ is the set of scalars, while the space of $1$-vectors $\R_m^{(1)}$ is isomorphic to $\R^m$ via the identification $\m{v}=(v_1,\ldots, v_m)^T \fd \sum_{j=1}^m v_j e_j$. When necessary, we shall use this interpretation of a column vector in $\R^m$ as a $1$-vector in $\R_m^{(1)}$.

An important automorphism on $\R_m$ leaving the multivector structure invariant is the {\em Clifford conjugation} $\ba{\cdot}$, which is defined as the linear mapping satisfying
\[
\ba{ab}=\ba{b}\ba{a}, \;\;\; a,b\in\R_m,  \;\;\;\;\;\; \mbox{and} \;\; \;\;\;\; \ba{e_j}=-e_j, \;\; j\in M.
\]
This leads to the norm on the Clifford algebra defined by $\|a\|^2=[a\ba{a}]_0=\sum_{A\subseteq M} a_A^2$ for $a\in\R_m$.

The Clifford product of two vectors  $\m{v},\m{u}\in \mathbb{R}_m^{(1)}$ may be written as $\m{v}\,\m{u}=\m{v}\cdot \m{u}+ \m{v}\wedge\m{u}$
where 
\begin{align*}
\m{v}\cdot \m{u} &= \frac{1}{2}(\m{v}\,\m{u}+ \m{u}\,\m{v}) = -\sum_{j=1}^m v_ju_j,  & & \mbox{and} & \m{v}\wedge \m{u} &= \frac{1}{2}(\m{v}\,\m{u}- \m{u}\,\m{v}) = \sum_{j<\ell} (v_ju_\el-v_\el u_j) e_j e_\el,
\end{align*}
are the so-called {\em dot} and {\em wedge} products respectively. Note that $\m{v}\cdot \m{u}=-\langle\m{v},\m{u}\rangle \in \R$, while $ \m{v}\wedge \m{u}\in \R_m^{(2)}$. In particular, for $\m{u}=\m{v}$ one obtains $\|\m{v}\|^2=-\m{v}^2$. Furthermore, for $\m{v}_1, \ldots, \m{v}_k\in \R_m^{(1)}$ we define the wedge (or exterior) product in terms of the Clifford product by
\[\m{v}_1\wedge \cdots \wedge \m{v}_k =  \frac{1}{k!} \sum_{\pi\in\textup{Sym}(k)}\textup{sgn}(\pi)\, \m{v}_{\pi(1)} \cdots \m{v}_{\pi(k)}  \; \in \; \R_m^{(k)}. \]
Among the most important properties of the wedge product we have
\begin{itemize}
\item  $\un{v}_{\pi(1)} \wedge \ldots \wedge \un{v}_{\pi(k)}= \text{sgn}(\pi) \un{v}_1 \wedge \ldots \wedge \un{v}_k$;
\item $\left[ \un{v}_1  \cdots \un{v}_k \right]_k = \un{v}_1 \wedge \ldots \wedge \un{v}_k $;
\item if $\{\un{v}_1, \ldots, \un{v}_k\}$ is a set of orthogonal vectors, then $\un{v}_1 \wedge \ldots \wedge \un{v}_k = \un{v}_1  \cdots \un{v}_k$;
\item $\un{v}_1 \wedge \ldots \wedge \un{v}_k=0$ if and only if  the vectors $\un{v}_1, \ldots, \un{v}_k$ are linearly dependent.
\end{itemize}

We shall also make use of the so-called {\it Gram matrix} of the vectors $\m{v}_1, \ldots, \m{v}_k\in \R^m$, which is defined as
\[G(\m{v}_1, \ldots, \m{v}_k)= \left(\begin{array}{cccc} \langle\m{v}_1, \m{v}_1\rangle &  \langle\m{v}_1, \m{v}_2\rangle & \ldots & \langle\m{v}_1, \m{v}_k\rangle \\[+.1cm]  
\langle\m{v}_2, \m{v}_1\rangle &  \langle\m{v}_2, \m{v}_2\rangle & \ldots & \langle\m{v}_2, \m{v}_k\rangle \\[+.1cm]  
\vdots & \vdots& \ddots & \vdots\\[+.1cm]
 \langle\m{v}_k, \m{v}_1\rangle &  \langle\m{v}_k, \m{v}_2\rangle & \ldots & \langle\m{v}_k, \m{v}_k\rangle \end{array}\right).\]
 The {\it Gram determinant} is the determinant of $G(\m{v}_1, \ldots, \m{v}_k)$ and can be expressed in terms of the wedge product of vectors by
 \begin{equation}\label{Gram_Det}
 \det G(\m{v}_1, \ldots, \m{v}_k) = \|  \un{v}_1 \wedge \ldots \wedge \un{v}_k\|^2.
 \end{equation}
 This determinant coincides with the square of the volume of the parallelepiped spanned by the vectors $\m{v}_1, \ldots, \m{v}_k$.
 
The most important operator in the theory of Clifford-valued functions is  the so-called Dirac operator (or gradient) defined as
\[
\partial_{\m{x}}=e_1\partial_{x_1}+\cdots+e_m\partial_{x_m}\, .
\]
The function theory centred around the null solutions of $\partial_{\m{x}}$ constitutes a natural and successful extension of classical complex analysis to higher dimensions. As the Dirac operator factorizes the Laplace operator
\[\Del_{\m{x}}=-\pa_{\m{x}}^2\, ,\]
this theory is also a refinement of harmonic analysis. Standard references on this setting are \cite{MR1130821, MR697564, MR1169463}.

\subsection{A distributional approach to integration} \label{CDDSec}
The central idea in the proofs of most of our Pizzetti formulae relies in the use of the concentrated delta distribution to describe integration on submanifolds of $\R^m$. Therefore, we now recall some basic notions related to this distribution, see \cite[Ch.~3]{MR0166596} for more details.

Let us consider  an $(m-k)$-surface $\Sigma\inc \R^m$  defined by means of $k$  equations of the form
\begin{equation}\label{PHfD}
\fhi_1(x_1, \ldots, x_m)=0, \hspace{.7cm} \fhi_2(x_1, \ldots, x_m)=0, \;\;\ldots,  \;\; \fhi_k(x_1, \ldots, x_m)=0,
\end{equation}
where the so-called {\it defining phase functions} $\fhi_1, \ldots, \fhi_k\in C^\infty(\R^m)$ are {\it independent}, i.e.\ 
\[\pa_{\m{x}}[\fhi_1]\wedge \ldots\ \wedge \pa_{\m{x}}[\fhi_k]\neq 0\;\;\; \mbox{ on } \;\;\; \Sigma\, ,\]
or equivalently, the gradients $\pa_{\m{x}}[\fhi_1], \ldots, \pa_{\m{x}}[\fhi_k]$ are linearly independent on every point of $\Sigma$.

The previous condition means that, at any point of $\Sigma$, there is a $k$-blade orthogonal to $\Sigma$ and therefore a ($m-k$)-dimensional tangent plane. We thus have have that, in an $m$-dimensional neighborhood $U$ of any point of $\Sigma$, there exists a $C^\infty$-local coordinate system  in which  the first $k$ coordinates are $u_1=\fhi_1$, $\ldots$, $u_k=\fhi_k$, and the remaining $u_{k+1}, \ldots, u_m$ can be chosen so that $J\hspace{-.1cm}\left(\stackanchor{\m{x}}{\m{u}}\right)>0$. Then, for any {\em test function} $\phi \in C^\infty(\R^m)$ with support in $U$, i.e.\ $\textup{supp} \, \phi \inc U$, one has
\[
\int_{\R^m} \del(\fhi_1)\cdots \del(\fhi_k)\phi(\m{x})\, dV_{\m{x}}=  \int_{\R^{m}} \del(u_1)\cdots \del(u_k) \, \psi(\m{u}) \, du_1 \cdots du_m,
\]
where $\psi(\m{u})= \phi(\m{x}(\m{u}))J\hspace{-.1cm}\left(\stackanchor{\m{x}}{\m{u}}\right)$ and $dV_{\m{x}}=dx_1\cdots dx_m$ is the classical Lebesgue measure in $\R^m$. It is thus natural to define $\del(\fhi_1)\cdots \del(\fhi_k)$ as 
\begin{equation}\label{del(P)_Def1}
(\del(\fhi_1)\ldots \del(\fhi_k), \phi ) = \int_{\R^m} \del(\fhi_1)\cdots \del(\fhi_k)\phi(\m{x})\, dV_{\m{x}}=\int \psi(0,\ldots, 0, u_{k+1},\ldots, u_m) \, du_{k+1} \cdots du_m.
\end{equation}
This last integral is taken over the surface $\Sigma\cap U$, which is why the {\em generalized function} $\del(\fhi_1)\ldots \del(\fhi_k)$ is said to be concentrated on this surface. This definition is extended to smooth functions $\phi$ with compact support on $\Sigma$, i.e.\ $\textup{supp} \, \phi \cap \Sigma$ compact, by considering a finite sum of integrals (\ref{del(P)_Def1}) over local parametrizations using partitions of unity, see e.g.\ \cite[Ch.\ 6]{MR2723362}.

In \cite{Ali_Pizz}, the following result was obtained to describe non-oriented integration of functions over the $(m-k)$-dimensional surface $\Sigma=\{\m{x}\in\R^m: \fhi_1(\m{x})=\ldots=\fhi_k(\m{x})=0\}$ in terms of the  generalized function $\del(\fhi_1)\ldots \del(\fhi_k)$. Let us consider an open region $\Om\inc \R^m$, then the result reads as follows.

\begin{teo}\label{Fund-Teo}
Let $\Sigma\inc \Om$ be a ($m-k$)-surface defined as in (\ref{PHfD}) by means of the independent phase functions  $\fhi_1, \ldots, \fhi_k\in C^\infty(\R^m)$. Then for any function $f \in C^\infty (\Om)$, with $\textup{supp}\, f \cap \Sigma$ compact, we have
\begin{equation}\label{NO_Int_For}
\int_\Sigma f\, dS= \int_{\R^m} \del(\fhi_1)\ldots \del(\fhi_k) \left\|\pa_{\m{x}}[\fhi_1]\wedge \ldots\ \wedge \pa_{\m{x}}[\fhi_k]\right\|f \, dV. 
\end{equation}
where $dS$ is the $(m-k)$-dimensional Lebesgue measure on $\Sigma$. 
\end{teo}
\begin{remark}
Theorem \ref{Fund-Teo} is a generalization to higher co-dimensions of Theorem 6.1.5 in \cite{MR1996773}.
\end{remark}

The above distributional approach can be easily adapted if we want to integrate over suitable regions of the ($m-k$)-surface $\Sigma$. Indeed, let $C= \{\m{x}\in\R^m:\fhi(\m{x})\leq 0\}$ be an $m$-dimensional region of $\R^m$ with $\fhi\in C^\infty(\R^m)$ and $\pa_{\m{x}}[\fhi]\wedge\pa_{\m{x}}[\fhi_1]\wedge \ldots\ \wedge \pa_{\m{x}}[\fhi_k] \neq 0$ on $\Sigma\cap \pa C:= \{\m{x}\in\Sigma:\fhi(\m{x})= 0\}$. Hence the integral over the region $\Sigma\cap C$  can be written as
\begin{equation}\label{NO_Int_For_Heav}
\int_{\Sigma\cap C} f\, dS= \int_{\R^m} H(-\fhi)\del(\fhi_1)\ldots \del(\fhi_k) \left\|\pa_{\m{x}}[\fhi_1]\wedge \ldots\ \wedge \pa_{\m{x}}[\fhi_k]\right\|f \, dV,
\end{equation}
where $H(-\fhi)=\begin{cases}1, & \fhi\leq0, \\ 0, & \fhi>0,\end{cases}$ is the Heaviside distribution.

As it is expected, the generalized function $\del(\fhi_1)\ldots \del(\fhi_k) \left\|\pa_{\m{x}}[\fhi_1]\wedge \ldots\ \wedge \pa_{\m{x}}[\fhi_k]\right\|$ is independent of the system of equations defining $\Sigma$.
To see this, let us transform the equations $\fhi_1=\ldots=\fhi_k=0$ to $\psi_1=\ldots=\psi_k=0$ where 
\[\psi_\el(\m{x})= \sum_{j=1}^k \al_{\el,j} (\m{x}) \fhi_j(\m{x})\, , \hspace{1cm} \el=1,\ldots, k\, .\]
Here the functions $\al_{\el,j}\in C^\infty(\R^m)$ are such that the matrix they form is nonsingular, i.e.\ $\det\{\al_{\el,j}\}\neq 0$ for every $\m{x}\in \R^m$. Obviously both sets of equations define the same manifold $\Sigma$. In these case, we have the following result (see \cite{Ali_Pizz}).
\begin{pro}\label{InvPro}
Let $\Sigma\inc \R^m$ be a ($m-k$)-surface defined by means of the independent phase functions  $\fhi_1, \ldots, \fhi_k\in C^\infty(\R^m)$ and let $\psi_\el= \sum_{j=1}^k \al_{\el,j}  \fhi_j$,  $\el=1,\ldots, k$, be new functions such that $\al_{\el,j}\in C^\infty(\R^m)$ and $\det\{\al_{\el,j}\}\neq 0$ for every $\m{x}\in \R^m$. We then have 
\[\del(\psi_1)\ldots \del(\psi_k) \, \left\|\pa_{\m{x}}[\psi_1]\wedge \ldots\ \wedge \pa_{\m{x}}[\psi_k]\right\|=  \del(\fhi_1)\ldots \del(\fhi_k) \,\left\|\pa_{\m{x}}[\fhi_1]\wedge \ldots\ \wedge \pa_{\m{x}}[\fhi_k]\right\|. \]
\end{pro}

\subsection{Integral over $\Sa^{m-1}$}
A good example to illustrate the invariance property established in Proposition \ref{InvPro} is the integral over the sphere $r \mathbb S^{m-1}:= \{\m{x}\in\R^m: \|\m{x}\|=r\}$, $r>0$. Indeed, $r\Sa^{m-1}$ can be defined by means of any of the two equations $\|\m{x}\|^2-r^2=0$ or $ \|\m{x}\|-r=0$. Both definitions can be used in (\ref{NO_Int_For}) without changing the result of the integral since
\[
\|\m{x}\|^2-r^2=\left( \|\m{x}\|+r\right) \left( \|\m{x}\|-r\right), \;\; \mbox{ and } \|\m{x}\|+r\neq 0, \mbox{ for all } \m{x}\in \R^m.
\]
Using these two phase functions defining  $\Sa^{m-1}$, and making use of the corresponding gradients 
\[
\pa_{\m{x}}\left[\|\m{x}\|^2-r^2\right]= 2\m{x}, \;\;\;\;\mbox{ and } \;\;\;\;\pa_{\m{x}}\left[\|\m{x}\|-r\right]=  \frac{\m{x}}{\|\m{x}\|},
\]
we obtain from  (\ref{NO_Int_For}) that 
\begin{equation}\label{DistIntSph}
\int_{r\Sa^{m-1}} f(\m{x}) dS_{\m{x}} = 2r \int_{\R^m} \del(\|\m{x}\|^2-r^2) f(\m{x}) dV_{\m{x}} = \int_{\R^m} \del(\|\m{x}\|-r) f(\m{x}) dV_{\m{x}}.
\end{equation}

Pizzetti's formula provides a method to compute integrals over the unit sphere $\Sa^{m-1}$ by acting with a certain power series of the Laplacian operator on the integrand, see \cite{Pizz}. For any polynomial $P:\R^m\fd \C$, this formula reads as 
\begin{equation}\label{PizSph}
\int_{\Sa^{m-1}} P(\m{x}) \; {d S_{\m{x}}} =  \sum_{k=0}^{\infty}  \frac{2 \pi^{m/2}}{2^{2k} k!\Gamma (k+m/2)} \Delta^k_{\m{x}} [P](0).
\end{equation}
From formula (\ref{PizSph}) we can easily derive the following Pizzetti-type formulae for the sphere $r \mathbb S^{m-1}$, and for the ball $B(0,r):=\{\m{x}\in\R^m: \|\m{x}\|<r\}$, with $r>0$.

\begin{pro}
Let $R$ be a polynomial in $\mathcal{P}(\R^m)$ and $r>0$. Then
\begin{align}
\int_{\mathbb S^{m-1}} \,R(r\underline{\omega}) \, dS_{\underline{\omega}} & = \sum_{k=0}^\infty \frac{2\pi^{\frac{m}{2}}}{2^{2k}k!\,\Gamma\left(\frac{m}{2}+k\right)} \, \Delta^k_{\m{x}} [R](0) \, r^{2k}, \label{Pi on Sph}\\[+0.2cm]
\int_{B(0,r)} \, R(\underline{x})\, d V_{\underline{x}}&=\sum_{k=0}^\infty\frac{\pi^{\frac{m}{2}}}{2^{2k}k!\,\Gamma\left(\frac{m}{2}+1+k\right)}\Delta^k_{\m{x}}[R](0)\, r^{2k+m}, \label{Pi on Ball} \\[+0.2cm]
\int_{r \mathbb S^{m-1}} \,R(\underline{x}) \, dS_{\underline{x}}&=\sum_{k=0}^\infty\frac{2\pi^{\frac{m}{2}}}{2^{2k}k!\,\Gamma\left(\frac{m}{2}+k\right)} \, \Delta^k_{\m{x}}[R](0) \, r^{2k+m-1}. \label{Pi on r-Sph}
\end{align}
\end{pro}
\pf 
Formula (\ref{Pi on Sph}) directly follows from (\ref{PizSph}) and from the identity $\Del_{\m{x}}\left[R(r\m{x})\right]= r^2 \Del_{\m{x}}\left[R\right](r\m{x})$. 
To prove formula (\ref{Pi on Ball}), we use spherical coordinates, i.e.\ $\m{x}=t\m{\omega}$ with $t=\|\m{x}\|$ and $\m{\omega}\in \Sa^{m-1}$. Hence
\begin{align*}
\int_{B(0,r)} \, R(\underline{x})\, d V_{\underline{x}}&=\int_0^r \int_{\mathbb S^{m-1}} R(t \underline{\omega}) \,  t^{m-1} \, dS_{\m{\omega}} \, dt\\
&= \int_0^r t^{m-1} \left( \int_{\mathbb S^{m-1}} R(t \underline{\omega}) \, dS_{\m{\omega}}\right) dt\\
&=\sum_{k=0}^\infty\frac{2\pi^{\frac{m}{2}}}{2^{2k}k!\,\Gamma(\frac{m}{2}+k)}\Delta_{\m{x}}^k[R](0) \left( \int_0^r t^{2k+m-1} \, dt \right)\\
&= \sum_{k=0}^\infty\frac{\pi^{\frac{m}{2}}}{2^{2k}k!\,\Gamma\left(\frac{m}{2}+1+k\right)}\Delta^k_{\m{x}}[R](0)\, r^{2k+m}.
\end{align*}
Finally, formula (\ref{Pi on r-Sph}) directly follows from the change of coordinates $\m{x}=r\m{\omega}$,  $\m{\omega}\in \Sa^{m-1}$, which implies that $ dS_{\m{x}}=r^{m-1}dS_{\m{\omega}}$. $\hfill\square$

\section{Pizzetti Formulae on regions of the sphere}
\label{Pizzetti on regions of sphere}
In this section we prove a handful of Pizetti-type formulae for integration of certain submanifolds of the unit sphere $\Sa^{m-1}$ and the unit ball $B(0,1)$. In particular, we are interested on the sub-spheres, sub-balls and spherical caps that arise when one intersects $\Sa^{m-1}$ and $B(0,1)$ with hyperplanes on $\R^{m}$ (see figures below). In order to write the integrals over these regions as the actions of the corresponding invariant differential operators, we will see that the language offered by the concentrated $\delta$ distribution becomes quite useful.

The Pizzetti-type formulae to be obtained in this section are summarized in the following theorem. 
\begin{teo}\label{PizzMainTheo}
Given $\m{\omega}\in\Sa^{m-1}$ and $p\in\R$, consider the hyperplane $H_{\m{\omega},p}:=\{\m{x}\in\R^m: \langle \m{x},\m{\omega}\rangle=p\}$.  Then following statements hold for $0\leq p<1$ and $R\in\mathcal{P}(\R^m)$.
\begin{itemize}
\item[$i)$] Let $\mathbb S_{p,\underline{\omega}}= \Sa^{m-1}\cap H_{\m{\omega},p}$ be the sub-sphere of $\Sa^{m-1}$ contained in the hyperplane $H_{\m{\omega},p}$. Then
\begin{equation}\label{Pi on S}
\int_{\mathbb S_{p,\underline{\omega}}} R(\underline{x})\, d S_{\underline{x}}=\sum_{k=0}^\infty\frac{2\pi^{\frac{m-1}{2}}}{2^{2k}k!\,\Gamma(\frac{m-1}{2}+k)} \left(\Delta_{\underline{x}}-{\langle \underline{\omega},\pa_{\underline{x}}\rangle}^2\right)^k [R] \Big|_{\underline{x}=p\underline{\omega}} \,(1-p^2)^{k+\frac{m}{2}-1}.
\end{equation}
\item[$ii)$] Let $\mathbb B_{p,\underline{\omega}}=B(0,1) \cap H_{\m{\omega},p}$ be the sub-ball of $B(0,1)$ contained in the hyperplane $H_{\m{\omega},p}$. Then
\begin{equation}\label{Pi on Hyp}
\int_{\mathbb B_{p,\underline{\omega}}} R(\underline{x}) \, d S_{\underline{x}}
=\sum_{k=0}^\infty\frac{\pi^{\frac{m-1}{2}}}{2^{2k}k!\,\Gamma(\frac{m+1}{2}+k)}\left(\Delta_{\underline{x}}-{\langle \underline{\omega},\pa_{\underline{x}}\rangle}^2\right)^k [R] \Big|_{\underline{x}=p\underline{\omega}} (1-p^2)^{k+\frac{m-1}{2}}.
\end{equation}
\item[$iii)$] Let $C_{p,\underline{\omega}}=\{\underline{x}\in\mathbb \Sa^{m-1}: \, \langle \underline{x},\underline{\omega}\rangle >p\}$ be the spherical cap in $\Sa^{m-1}$  located in the upper-half of $\R^{m+1}$ relative to $H_{\m{\omega},p}$. Then, for $R_\el\in\mathcal{P}_\el(\R^m)$ one has,
\begin{equation}\label{Pi on Cap Hom}
\int_{C_{p,\underline{\omega}}} R_\el(\underline{x}) \, d S_{\underline{x}}=\sum_{k=0}^{\lfloor \frac{\el}{2}\rfloor}\frac{2\pi^{\frac{m-1}{2}}}{2^{2k}k!\,\Gamma(\frac{m-1}{2}+k)} c_{k, \el}(p)
\left(\Delta_{\underline{x}}-{\langle \underline{\omega},\pa_{\underline{x}}\rangle}^2\right)^k[R_\el] \Big|_{\underline{x}=\underline{\omega}} ,
\end{equation}
where the coefficients $c_{k, \el}(p)$ are given by
\begin{align*}
c_{k, \el}(p) &= \int_p^1 y_1^{\el-2k}(1-y_1^2)^{k+\frac{m-3}{2}}\,dy_1 \\
&= \frac{\Gamma\left(\frac{\el+1}{2}-k\right)\Gamma\left(k+ \frac{m-1}{2}\right)}{2\Gamma\left( \frac{\el+m}{2}\right)} - \frac{p^{\el-2k+1}}{\el-2k+1} {}_2F_1\left(-k-\frac{m-3}{2}, \frac{\el+1}{2}-k;\frac{\el+3}{2}-k, p^2\right).
\end{align*}
\item[$iv)$] Similarly, we obtain a Pizzetti formula for the spherical cap located in the lower-half of $\R^{m+1}$ relative to $H_{\m{\omega},p}$, i.e.\ $C'_{p,\m{w}}:=\{\underline{x}\in\mathbb \Sa^{m-1}: \, \langle \underline{x},\underline{\omega}\rangle <p\}$. In this case we obtain for  $R_\el\in\mathcal{P}_\el(\R^m)$ that 
\begin{equation}\label{Pi on Cap Hom Comp}
\int_{C'_{p,\underline{\omega}}} R_\el(\underline{x}) \, d S_{\underline{x}}=\sum_{k=0}^{\lfloor \frac{\el}{2}\rfloor}\frac{2\pi^{\frac{m-1}{2}}}{2^{2k}k!\,\Gamma(\frac{m-1}{2}+k)} c'_{k, \el}(p)
\left(\Delta_{\underline{x}}-{\langle \underline{\omega},\pa_{\underline{x}}\rangle}^2\right)^k[R_\el] \Big|_{\underline{x}=\underline{\omega}} ,
\end{equation}
where the coefficients $c'_{k, \el}(p)$ are given by
\begin{align*}
c'_{k, \el}(p) &= \int_{-1}^p y_1^{\el-2k}(1-y_1^2)^{k+\frac{m-1}{2}}\,dy_1 \\
&=\frac{p^{\el-2k+1}}{\el-2k+1} {}_2F_1\left(-k-\frac{m-3}{2}, \frac{\el+1}{2}-k;\frac{\el+3}{2}-k, p^2\right) + (-1)^\el  \frac{\Gamma\left(\frac{\el+1}{2}-k\right)\Gamma\left(k+ \frac{m-1}{2}\right)}{2\Gamma\left( \frac{\el+m}{2}\right)}.
\end{align*}
\end{itemize}
\end{teo}

{In Figure 1, we depict (for the case $m=3$) the intersection of the sphere $\Sa^{m-1}$ with the hyperplane  $H_{p, \m{\omega}}$. The blue contour represents the subsphere $\mathbb S_{p,\underline{\omega}}$ in part $i)$ of the theorem, while the red shaded area represents the sub-ball  $\mathbb B_{p,\underline{\omega}}$ in part $ii)$.}
\begin{center}
\begin{figure}[H]
           \centering
            \resizebox{0.3\textwidth}{!}{
            \tikzset{every picture/.style={line width=0.75pt}} 

\begin{tikzpicture}[x=0.75pt,y=0.75pt,yscale=-1,xscale=1]

\draw   (30,137.65) .. controls (30,62.18) and (91.21,1) .. (166.72,1) .. controls (242.22,1) and (303.43,62.18) .. (303.43,137.65) .. controls (303.43,213.12) and (242.22,274.3) .. (166.72,274.3) .. controls (91.21,274.3) and (30,213.12) .. (30,137.65) -- cycle ;
\draw  [dash pattern={on 0.84pt off 2.51pt}]  (30,137.65) .. controls (73.52,102.85) and (140.33,89.97) .. (199.41,96.07) .. controls (241.94,100.47) and (280.46,114.69) .. (303.43,137.65) ;
\draw    (30,137.65) .. controls (82.05,173.46) and (157.92,185.01) .. (222.55,172.31) .. controls (253.32,166.25) and (281.54,154.7) .. (303.43,137.65) ;
\draw    (166.72,137.65) -- (301.43,137.65) ;
\draw [shift={(303.43,137.65)}, rotate = 180] [color={rgb, 255:red, 0; green, 0; blue, 0 }  ][line width=0.75]    (10.93,-4.9) .. controls (6.95,-2.3) and (3.31,-0.67) .. (0,0) .. controls (3.31,0.67) and (6.95,2.3) .. (10.93,4.9)   ;
\draw [shift={(166.72,137.65)}, rotate = 0] [color={rgb, 255:red, 0; green, 0; blue, 0 }  ][fill={rgb, 255:red, 0; green, 0; blue, 0 }  ][line width=0.75]      (0, 0) circle [x radius= 3.35, y radius= 3.35]   ;
\draw    (243.42,26.39) -- (166.72,137.65) ;
\draw [shift={(166.72,137.65)}, rotate = 124.58] [color={rgb, 255:red, 0; green, 0; blue, 0 }  ][fill={rgb, 255:red, 0; green, 0; blue, 0 }  ][line width=0.75]      (0, 0) circle [x radius= 3.35, y radius= 3.35]   ;
\draw [shift={(244.55,24.74)}, rotate = 124.58] [color={rgb, 255:red, 0; green, 0; blue, 0 }  ][line width=0.75]    (10.93,-4.9) .. controls (6.95,-2.3) and (3.31,-0.67) .. (0,0) .. controls (3.31,0.67) and (6.95,2.3) .. (10.93,4.9)   ;
\draw [color={rgb, 255:red, 46; green, 107; blue, 178 }  ,draw opacity=1 ][line width=1.5]    (244.55,24.74) .. controls (232.71,41.81) and (224.06,61.99) .. (218.62,83.59) .. controls (204.36,140.25) and (212.21,206.66) .. (242.56,252.16) ;
\draw [color={rgb, 255:red, 74; green, 144; blue, 226 }  ,draw opacity=1 ][line width=1.5]  [dash pattern={on 1.69pt off 2.76pt}]  (244.55,24.74) .. controls (281.48,87.58) and (287.46,178.35) .. (242.56,252.16) ;
\draw [color={rgb, 255:red, 208; green, 2; blue, 27 }  ,draw opacity=1 ][fill={rgb, 255:red, 171; green, 98; blue, 98 }  ,fill opacity=1 ] [dash pattern={on 4.5pt off 4.5pt}]  (274.49,115.71) -- (213.62,173.36) ;
\draw [color={rgb, 255:red, 208; green, 2; blue, 27 }  ,draw opacity=1 ] [dash pattern={on 4.5pt off 4.5pt}]  (270.5,106.53) -- (213.62,161.39) ;
\draw [color={rgb, 255:red, 208; green, 2; blue, 27 }  ,draw opacity=1 ] [dash pattern={on 4.5pt off 4.5pt}]  (274.49,126.48) -- (217.61,181.34) ;
\draw [color={rgb, 255:red, 208; green, 2; blue, 27 }  ,draw opacity=1 ] [dash pattern={on 4.5pt off 4.5pt}]  (274.49,137.45) -- (217.61,192.31) ;
\draw [color={rgb, 255:red, 208; green, 2; blue, 27 }  ,draw opacity=1 ] [dash pattern={on 4.5pt off 4.5pt}]  (271.5,151.41) -- (219.61,202.28) ;
\draw [color={rgb, 255:red, 208; green, 2; blue, 27 }  ,draw opacity=1 ] [dash pattern={on 4.5pt off 4.5pt}]  (271.5,164.38) -- (220.6,214.25) ;
\draw [color={rgb, 255:red, 208; green, 2; blue, 27 }  ,draw opacity=1 ] [dash pattern={on 4.5pt off 4.5pt}]  (267.51,185.33) -- (226.59,222.23) ;
\draw [color={rgb, 255:red, 208; green, 2; blue, 27 }  ,draw opacity=1 ] [dash pattern={on 4.5pt off 4.5pt}]  (268.5,95.56) -- (211.62,150.42) ;
\draw [color={rgb, 255:red, 208; green, 2; blue, 27 }  ,draw opacity=1 ] [dash pattern={on 4.5pt off 4.5pt}]  (266.51,84.59) -- (209.63,139.45) ;
\draw [color={rgb, 255:red, 208; green, 2; blue, 27 }  ,draw opacity=1 ] [dash pattern={on 4.5pt off 4.5pt}]  (264.51,73.61) -- (212.62,123.49) ;
\draw [color={rgb, 255:red, 208; green, 2; blue, 27 }  ,draw opacity=1 ] [dash pattern={on 4.5pt off 4.5pt}]  (263.51,63.64) -- (212.62,111.52) ;
\draw [color={rgb, 255:red, 208; green, 2; blue, 27 }  ,draw opacity=1 ] [dash pattern={on 4.5pt off 4.5pt}]  (260.52,56.66) -- (212.62,101.54) ;
\draw [color={rgb, 255:red, 208; green, 2; blue, 27 }  ,draw opacity=1 ] [dash pattern={on 4.5pt off 4.5pt}]  (256.53,50.67) -- (216.61,88.58) ;
\draw [color={rgb, 255:red, 208; green, 2; blue, 27 }  ,draw opacity=1 ] [dash pattern={on 4.5pt off 4.5pt}]  (253.54,43.69) -- (221.6,73.61) ;
\draw [color={rgb, 255:red, 208; green, 2; blue, 27 }  ,draw opacity=1 ] [dash pattern={on 4.5pt off 4.5pt}]  (250.54,36.71) -- (227.39,57.26) ;
\draw [color={rgb, 255:red, 208; green, 2; blue, 27 }  ,draw opacity=1 ] [dash pattern={on 4.5pt off 4.5pt}]  (263.51,199.29) -- (231.58,229.22) ;
\draw [color={rgb, 255:red, 208; green, 2; blue, 27 }  ,draw opacity=1 ] [dash pattern={on 4.5pt off 4.5pt}]  (259.52,217.25) -- (235.57,238.19) ;
\draw [color={rgb, 255:red, 208; green, 2; blue, 27 }  ,draw opacity=1 ] [dash pattern={on 4.5pt off 4.5pt}]  (254.53,231.21) -- (236.57,247.17) ;
\draw    (249.54,137.65) ;
\draw [shift={(249.54,137.65)}, rotate = 0] [color={rgb, 255:red, 0; green, 0; blue, 0 }  ][fill={rgb, 255:red, 0; green, 0; blue, 0 }  ][line width=0.75]      (0, 0) circle [x radius= 3.35, y radius= 3.35]   ;

\draw (207.62,40.28) node [anchor=north west][inner sep=0.75pt]    {$\underline{x}$};
\draw (311.4,125.06) node [anchor=north west][inner sep=0.75pt]    {$\underline{\omega }$};
\draw (199.1,141.04) node [anchor=north west][inner sep=0.75pt]  [rotate=-0.3]  {$p$};

\end{tikzpicture}  
           }
           \caption{Intersection of the sphere with a hyperplane}
\end{figure}
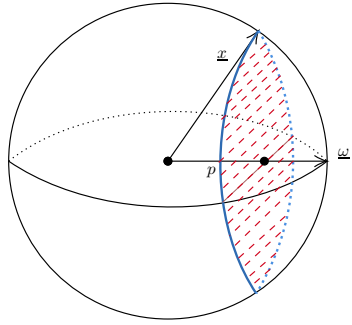    
\end{center}
{Similarly, Figure 2 depicts the two complementary caps $C_{p,\underline{\omega}}$ and $C'_{p,\m{w}}$ in purple and green, respectively.}
\begin{center}
      \begin{figure}[H]
           \centering
            \resizebox{0.3\textwidth}{!}{
            \tikzset{every picture/.style={line width=0.75pt}} 

\begin{tikzpicture}[x=0.75pt,y=0.75pt,yscale=-1,xscale=1]

\draw    (168,137.2) -- (303,137.2) ;
\draw [shift={(305,137.2)}, rotate = 180] [color={rgb, 255:red, 0; green, 0; blue, 0 }  ][line width=0.75]    (10.93,-4.9) .. controls (6.95,-2.3) and (3.31,-0.67) .. (0,0) .. controls (3.31,0.67) and (6.95,2.3) .. (10.93,4.9)   ;
\draw [shift={(168,137.2)}, rotate = 0] [color={rgb, 255:red, 0; green, 0; blue, 0 }  ][fill={rgb, 255:red, 0; green, 0; blue, 0 }  ][line width=0.75]      (0, 0) circle [x radius= 3.35, y radius= 3.35]   ;
\draw    (244.87,25.65) -- (168,137.2) ;
\draw [shift={(168,137.2)}, rotate = 124.57] [color={rgb, 255:red, 0; green, 0; blue, 0 }  ][fill={rgb, 255:red, 0; green, 0; blue, 0 }  ][line width=0.75]      (0, 0) circle [x radius= 3.35, y radius= 3.35]   ;
\draw [shift={(246,24)}, rotate = 124.57] [color={rgb, 255:red, 0; green, 0; blue, 0 }  ][line width=0.75]    (10.93,-4.9) .. controls (6.95,-2.3) and (3.31,-0.67) .. (0,0) .. controls (3.31,0.67) and (6.95,2.3) .. (10.93,4.9)   ;
\draw [color={rgb, 255:red, 0; green, 0; blue, 0 }  ,draw opacity=1 ][line width=1.5]    (246,24) .. controls (234.13,41.11) and (225.46,61.35) .. (220.01,83) .. controls (205.72,139.81) and (213.59,206.39) .. (244,252) ;
\draw [color={rgb, 255:red, 0; green, 0; blue, 0 }  ,draw opacity=1 ][line width=1.5]  [dash pattern={on 1.69pt off 2.76pt}]  (246,24) .. controls (283,87) and (289,178) .. (244,252) ;
\draw    (251,137.2) ;
\draw [shift={(251,137.2)}, rotate = 0] [color={rgb, 255:red, 0; green, 0; blue, 0 }  ][fill={rgb, 255:red, 0; green, 0; blue, 0 }  ][line width=0.75]      (0, 0) circle [x radius= 3.35, y radius= 3.35]   ;
\draw [color={rgb, 255:red, 85; green, 141; blue, 23 }  ,draw opacity=1 ][line width=1.5]    (31,137.2) .. controls (78,172.2) and (158,183.2) .. (215,173.2) ;
\draw [color={rgb, 255:red, 85; green, 141; blue, 23 }  ,draw opacity=1 ][line width=1.5]    (246,24) .. controls (147,-38.8) and (33,31.2) .. (31,137.2) ;
\draw [color={rgb, 255:red, 85; green, 141; blue, 23 }  ,draw opacity=1 ][line width=1.5]    (31,137.2) .. controls (33,255.2) and (160,307.2) .. (244,252) ;
\draw [color={rgb, 255:red, 85; green, 141; blue, 23 }  ,draw opacity=1 ][line width=1.5]  [dash pattern={on 1.69pt off 2.76pt}]  (31,137.2) .. controls (104,82.2) and (210,85.2) .. (274,115.2) ;
\draw [color={rgb, 255:red, 189; green, 16; blue, 224 }  ,draw opacity=1 ][line width=1.5]  [dash pattern={on 1.69pt off 2.76pt}]  (274,115.2) .. controls (274,114.64) and (282.14,119.4) .. (290.18,124.93) .. controls (296.52,129.29) and (302.8,134.12) .. (305,137.2) ;
\draw [color={rgb, 255:red, 189; green, 16; blue, 224 }  ,draw opacity=1 ][line width=1.5]    (215,173.2) .. controls (229.4,171.4) and (242.25,168.17) .. (253.82,164.08) .. controls (266.64,159.55) and (277.9,153.95) .. (287.97,148.05) .. controls (294.05,144.49) and (299.7,140.81) .. (305,137.2) ;
\draw [color={rgb, 255:red, 189; green, 16; blue, 224 }  ,draw opacity=1 ][line width=1.5]    (244,252) .. controls (296.56,211.37) and (312.85,157.41) .. (302.54,109.28) .. controls (295.4,75.93) and (275.48,45.37) .. (246,24) ;
\draw [color={rgb, 255:red, 0; green, 0; blue, 0 }  ,draw opacity=1 ][fill={rgb, 255:red, 171; green, 98; blue, 98 }  ,fill opacity=1 ] [dash pattern={on 3.75pt off 3pt on 7.5pt off 1.5pt}]  (278,118.2) -- (217,176) ;
\draw [color={rgb, 255:red, 0; green, 0; blue, 0 }  ,draw opacity=1 ] [dash pattern={on 3.75pt off 3pt on 7.5pt off 1.5pt}]  (274,109) -- (217,164) ;
\draw [color={rgb, 255:red, 0; green, 0; blue, 0 }  ,draw opacity=1 ] [dash pattern={on 3.75pt off 3pt on 7.5pt off 1.5pt}]  (278,129) -- (221,184) ;
\draw [color={rgb, 255:red, 0; green, 0; blue, 0 }  ,draw opacity=1 ] [dash pattern={on 3.75pt off 3pt on 7.5pt off 1.5pt}]  (278,140) -- (221,195) ;
\draw [color={rgb, 255:red, 0; green, 0; blue, 0 }  ,draw opacity=1 ] [dash pattern={on 3.75pt off 3pt on 7.5pt off 1.5pt}]  (274,156) -- (222,207) ;
\draw [color={rgb, 255:red, 0; green, 0; blue, 0 }  ,draw opacity=1 ] [dash pattern={on 3.75pt off 3pt on 7.5pt off 1.5pt}]  (273,170) -- (226,214.2) ;
\draw [color={rgb, 255:red, 0; green, 0; blue, 0 }  ,draw opacity=1 ] [dash pattern={on 3.75pt off 3pt on 7.5pt off 1.5pt}]  (270,186) -- (229,223) ;
\draw [color={rgb, 255:red, 0; green, 0; blue, 0 }  ,draw opacity=1 ] [dash pattern={on 3.75pt off 3pt on 7.5pt off 1.5pt}]  (272,98) -- (215,153) ;
\draw [color={rgb, 255:red, 0; green, 0; blue, 0 }  ,draw opacity=1 ] [dash pattern={on 3.75pt off 3pt on 7.5pt off 1.5pt}]  (270,87) -- (213,142) ;
\draw [color={rgb, 255:red, 0; green, 0; blue, 0 }  ,draw opacity=1 ] [dash pattern={on 3.75pt off 3pt on 7.5pt off 1.5pt}]  (268,76) -- (216,126) ;
\draw [color={rgb, 255:red, 0; green, 0; blue, 0 }  ,draw opacity=1 ] [dash pattern={on 3.75pt off 3pt on 7.5pt off 1.5pt}]  (267,66) -- (216,114) ;
\draw [color={rgb, 255:red, 0; green, 0; blue, 0 }  ,draw opacity=1 ] [dash pattern={on 3.75pt off 3pt on 7.5pt off 1.5pt}]  (264,59) -- (216,104) ;
\draw [color={rgb, 255:red, 0; green, 0; blue, 0 }  ,draw opacity=1 ] [dash pattern={on 3.75pt off 3pt on 7.5pt off 1.5pt}]  (260,53) -- (220,91) ;
\draw [color={rgb, 255:red, 0; green, 0; blue, 0 }  ,draw opacity=1 ] [dash pattern={on 3.75pt off 3pt on 7.5pt off 1.5pt}]  (257,46) -- (225,76) ;
\draw [color={rgb, 255:red, 0; green, 0; blue, 0 }  ,draw opacity=1 ] [dash pattern={on 3.75pt off 3pt on 7.5pt off 1.5pt}]  (254,39) -- (230.8,59.6) ;
\draw [color={rgb, 255:red, 0; green, 0; blue, 0 }  ,draw opacity=1 ] [dash pattern={on 3.75pt off 3pt on 7.5pt off 1.5pt}]  (265,204) -- (233,234) ;
\draw [color={rgb, 255:red, 0; green, 0; blue, 0 }  ,draw opacity=1 ] [dash pattern={on 3.75pt off 3pt on 7.5pt off 1.5pt}]  (260,220) -- (236,241) ;

\draw (209,39.6) node [anchor=north west][inner sep=0.75pt]    {$\underline{x}$};
\draw (312,123.6) node [anchor=north west][inner sep=0.75pt]    {$\underline{\omega }$};
\draw (200.47,140.62) node [anchor=north west][inner sep=0.75pt]  [rotate=-0.3]  {$p$};

\end{tikzpicture}  
           }
          \caption{Spherical Caps}
\end{figure}
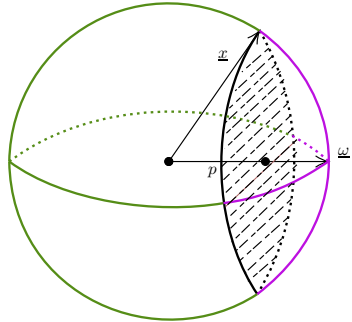     
\end{center}

We note that that the two formulae (\ref{Pi on Cap Hom}) and (\ref{Pi on Cap Hom Comp}) describe integration over complementary caps on $\Sa^{m-1}$, therefore their sum yields an alternative Pizzetti formula on $\Sa^{m-1}$. Indeed, from the sum of these two we obtain:
\begin{align}\label{PizzMod}
\int_{\Sa^{m-1}} R_\el(\underline{x}) \, d S_{\underline{x}}&=\sum_{k=0}^{\lfloor \frac{\el}{2}\rfloor}\frac{2\pi^{\frac{m-1}{2}}}{2^{2k}k!\,\Gamma(\frac{m-1}{2}+k)}    \left( \int_{-1}^1 y_1^{\el-2k}(1-y_1^2)^{k+\frac{m-1}{2}}\,dy_1 \right)
\left(\Delta_{\underline{x}}-{\langle \underline{\omega},\pa_{\underline{x}}\rangle}^2\right)^k[R_\el] \Big|_{\underline{x}=\underline{\omega}} \nonumber\\
&= \left((-1)^\el+1\right) \, \frac{\pi^{\frac{m-1}{2}}}{\Gamma\left( \frac{\el+m}{2}\right)} \, \sum_{k=0}^{\lfloor \frac{\el}{2}\rfloor}\frac{ \Gamma\left(\frac{\el+1}{2}-k\right)}{2^{2k}k!}
\left(\Delta_{\underline{x}}-{\langle \underline{\omega},\pa_{\underline{x}}\rangle}^2\right)^k[R_\el] \Big|_{\underline{x}=\underline{\omega}}.
\end{align}
If $\el$ is odd, it is clear that both sides of (\ref{PizzMod}) vanish. On the other hand, for $\el = 2s$, we obtain the following consequence of Theorem \ref{PizzMainTheo}.
\begin{cor}
Let  $R_{2s}\in\mathcal{P}_{2s}(\R^m)$. Then
\[
\int_{\Sa^{m-1}} R_{2s}(\underline{x}) \, d S_{\underline{x}}= \frac{2\pi^{\frac{m-1}{2}}}{\Gamma\left( s+\frac{m}{2}\right)} \, \sum_{k=0}^s \frac{ \Gamma\left(s-k + \frac{1}{2}\right)}{2^{2k}k!}
\left(\Delta_{\underline{x}}-{\langle \underline{\omega},\pa_{\underline{x}}\rangle}^2\right)^k[R_{2s}] \Big|_{\underline{x}=\underline{\omega}}, 
\]
where $\m{\omega}\in\Sa^{m-1}$ is an arbitrary unit vector. When compared with the classical Pizzetti formula (\ref{PizSph}) on $\Sa^{m-1}$, the above formula yields the identity
\begin{equation}\label{Ann}
\Del_{\m{x}}^s [R_{2s}] \Big|_{\underline{x}=0} = \frac{2^{2s} s!}{\pi^{\frac{1}{2}}} \sum_{k=0}^s \frac{ \Gamma\left(s-k + \frac{1}{2}\right)}{2^{2k}k!}
\left(\Delta_{\underline{x}}-{\langle \underline{\omega},\pa_{\underline{x}}\rangle}^2\right)^k[R_{2s}] \Big|_{\underline{x}=\underline{\omega}}.
\end{equation}
\end{cor}
{This} result shows that the expression on the right hand side of (\ref{Ann}) is independent of the unit vector $\m{\omega}$ and in Appendix \ref{App}, we provide a direct proof for this identity. Now, we proceed to prove the formulae in Theorem \ref{PizzMainTheo}.


\subsection{Integration on the subsphere $\mathbb S_{p,\underline{\omega}}$}
Our first goal is to integrate over the intersection $\mathbb S_{p,\underline{\omega}}$ of the unit sphere $\Sa^{m-1}$ with the hyperplane $H_{\m{\omega},p}$, i.e.:
\[
\mathbb S_{p,\underline{\omega}}=\{\underline{x}\in\mathbb R^n: \, \langle \underline{x},\underline{\omega}\rangle =p,\, \| \underline{x}\|=1\}, \;\;\;\;\;\;\; \m{\omega}\in\Sa^{m-1}, \;\;0\leq p<1.
\]
This is an $(m-2)$-dimensional sphere on the hyperplane $H_{\m{\omega},p}$  centered at 
$p\, \underline{\omega}$ and with radius $\displaystyle (1-p^2)^{\frac{1}{2}}$, {see Figure 1}. A simple computation shows that this is indeed the case
\[
\|\underline{x}-p\, \underline{\omega}\|^2=\langle \underline{x}-p\underline{\omega}, \underline{x}-p\underline{\omega} \rangle =\|\underline{x}\|^2-2p\, \langle \underline{x},\underline{\omega}\rangle +p^2 \|\m{\omega}\|^2=1-2p^2+p^2=1-p^2.
\]
The subsphere $\Sa_{p,\underline{\omega}}$ is defined by the pair of smooth functions $\fhi_1( \underline{x})=\|\underline{x}\|^2-1$ and $\fhi_2( \underline{x})= \langle \underline{x},\underline{\omega}\rangle -p$, whose gradients are given by $\partial_{\underline{x}}[\fhi_1]=2\underline{x}$ and $\partial_{\underline{x}}[\fhi_2]=\underline{\omega}$ respectively. We thus obtain
\[
\partial_{\underline{x}}[\fhi_1] \wedge \partial_{\underline{x}}[\fhi_2]=2\underline{x}\wedge \underline{\omega}=2( \underline{x} \, \underline{\omega}+\langle \underline{x},\underline{\omega}\rangle)=2( \underline{x} \, \underline{\omega}+p)\, ,
\]
which yields
\begin{equation}\label{wedge phi}
\|\partial_{\underline{x}}[\fhi_1] \wedge \partial_{\underline{x}}[\fhi_2]\|^2=4( \underline{x} \, \underline{\omega}+p)\overline{(\underline{x} \, \underline{\omega}+p)} = 4( \underline{x} \, \underline{\omega}+p) ( \underline{\omega} \, \underline{x}+p) = 4(1-p^2)\, .
\end{equation}

\begin{remark}
We can see this in a a more geometric way. Indeed, we can write $\underline{x} =\cos(\theta) \underline{\omega}+\sin(\theta) \underline{\epsilon}$ with $ \underline{\epsilon}\perp  \underline{\omega}$. Hence $\underline{x} \wedge \underline{\omega} =  \underline{\epsilon} \wedge \underline{\omega} \sin(\theta)=\underline{\epsilon}  \, \underline{\omega} \sin(\theta)$,
which implies $|\underline{x} \wedge \underline{\omega}| =|\sin(\theta)|=(1-\cos(\theta)^2)^{\frac{1}{2}}$ where $\cos(\theta)=\langle \underline{x},\underline{\omega}\rangle =p$.
\end{remark}

Using (\ref{wedge phi}) and Theorem \ref{Fund-Teo} we can now write the integral of a polynomial $R$ on $\mathbb S_{p,\underline{\omega}}$ as 
\begin{align*}
\int_{\mathbb S_{p,\underline{\omega}}} R(\underline{x})\, d S_{\underline{x}}&= \int_{\mathbb R^m} \del(\fhi_1)\del(\fhi_2) \|\partial_{\underline{x}}[\fhi_1] \wedge \partial_{\underline{x}}[\fhi_2]\| R(\underline{x})\, dV_{\underline{x}}\\
&= 2\sqrt{1-p^2} \int_{\mathbb R^m} \del(\|\underline{x}\|^2-1)\del(\langle \underline{x},\underline{\omega}\rangle-p) R(\underline{x})\, dV_{\underline{x}}.
\end{align*}
Let us now consider the change of coordinates $\underline{y}=M \underline{x}$, where $M\in \textup{SO}(m)$ is a rotation matrix whose first row is given by the unit vector $\m{\omega}\in\Sa^{m-1}$, i.e.\ the first component of $\m{y}$ is $y_1=\langle \underline{x},\underline{\omega}\rangle$. Under this transformation, the above integral transforms into:
\begin{align}\label{IntInt}
\int_{\mathbb S_{p,\underline{\omega}}} R(\underline{x})\, d S_{\underline{x}}&= 2\sqrt{1-p^2} \int_{\mathbb R^m} \del(\|\underline{y}\|^2-1)\del(y_1-p) R(M^{-1}\underline{y})\, dV_{\underline{y}}\nonumber\\
&= 2\sqrt{1-p^2} \int_{\mathbb R^{m-1}} \del(y_2^2+\dots +y_n^2-(1-p^2)) \, R(M^{-1}(p,y_2,\dots,y_m)^T)\; dy_2\dots dy_m.
\end{align}
The last expression corresponds to {an} integral over the sphere $\displaystyle (1-p^2)^{\frac{1}{2}}\mathbb S^{m-2}\subset \mathbb R^{m-1}$ with respect to the vector $(y_2,\dots,y_m)$, see (\ref{DistIntSph}). Thus, applying the Pizzetti formula (\ref{Pi on r-Sph}) on this sphere, we obtain:
\[
\int_{\mathbb S_{p,\underline{\omega}}} R(\underline{x})\, d S_{\underline{x}}=\sum_{k=0}^\infty\frac{2\pi^{\frac{m-1}{2}}}{2^{2k}k!\,\Gamma(\frac{m-1}{2}+k)}(\Delta_{\underline{y}}-\partial_{y_1}^2)^k R(M^{-1}(p,y_2,\dots,y_m)^T)\Big|_{y_2,\ldots,y_m=0} (1-p^2)^{k+\frac{m-2}{2}}.
\]
Returning to the original coordinate system $\m{x}=M^{-1} \m{y}$, and making use of the identities $\Del_{\m{y}}=\Del_{\m{x}}$ and $\pa_{y_1}=\langle\m{\omega}, \pa_{\m{x}}\rangle=\sum_{j=1}^m \omega_j\pa_{x_j}$, we have:
\begin{align*}
(\Delta_{\underline{y}}-\partial_{y_1}^2)^k R(M^{-1}(p,y_2,\dots,y_m)^T)\Big|_{y_2,\ldots,y_m=0}=\left(\Delta_{\underline{x}}-\langle\m{\omega}, \pa_{\m{x}}\rangle^2\right)^k [R]\Big|_{\underline{x}=M^{-1}(p, 0 \dots 0)^T}.
\end{align*}
Finally, observe that the point at which the last expression is evaluated is given by $\m{x}=p\m{\omega}$. Indeed, it is easily seen that $Mp\m{\omega}=(p, 0,\dots 0)^T$, and the Pizzetti formula for $\mathbb S_{p,\underline{\omega}}$ reads as:
\[
\int_{\mathbb S_{p,\underline{\omega}}} R(\underline{x})\, d S_{\underline{x}}=\sum_{k=0}^\infty\frac{2\pi^{\frac{m-1}{2}}}{2^{2k}k!\,\Gamma(\frac{m-1}{2}+k)}\left(\Delta_{\underline{x}}-\langle\m{\omega}, \pa_{\m{x}}\rangle^2\right)^k [R]\Big|_{\underline{x}=p\underline{\omega}} \, (1-p^2)^{k+\frac{m}{2}-1},
\]
which proves the first part of the Theorem \ref{PizzMainTheo} $i)$.

\subsection{Integration on the sub-ball $\mathbb B_{p,\underline{\omega}}$}
We now turn our attention to the integral over the intersection $\mathbb B_{p,\underline{\omega}}$ of the unit ball $B(0,1)$ with the hyperplane $H_{\m{\omega},p}$, i.e.\
\[
\mathbb B_{p,\underline{\omega}}=\{\underline{x}\in\mathbb R^n: \, \langle \underline{x},\underline{\omega}\rangle =p,\, \| \underline{x}\|<1\}, \;\;\;\;\;\;\;  \m{\omega}\in\Sa^{m-1}, \;\;0\leq p<1.
\]
Similarly to the previous case, $\mathbb B_{p,\underline{\omega}}$ is an $(m-1)$-dimensional ball on the hyperplane $H_{\m{\omega},p}$  centered at $p\, \underline{\omega}$ and with radius $\displaystyle (1-p^2)^{\frac{1}{2}}$, {see Figure 1}.  

From (\ref{NO_Int_For_Heav}) we obtain that
\begin{equation}
\label{Pi on B}
\int_{\mathbb B_{p,\underline{\omega}}} R(\underline{x}) \, d S_{\underline{x}}=\int_{\mathbb R^m} H(1-\|\underline{x}\|)
\delta({\langle \underline{x},\underline{\omega}\rangle}-p) R(\underline{x})\,dV_{\underline{x}}.
\end{equation}
Just as before, we consider a rotation $\underline{y}=M \underline{x}$ where the first row of matrix $M\in \textup{SO}(m)$ is given by $\m{\omega}\in\Sa^{m-1}$, i.e.\ $y_1=\langle \underline{x},\underline{\omega}\rangle$. With this change of coordinates in the above integral we obtain:
\begin{align*}
\int_{\mathbb B_{p,\underline{\omega}}} R(\underline{x}) \, d S_{\underline{x}}&=\int_{\mathbb R^m} H(1-\|\underline{y}\|)
\delta(y_1-p) R(M^{-1}\underline{y})\,dV_{\underline{x}}\\
&=\int_{\mathbb R^{m-1}} H(1-p^2-(y_2^2+\dots y_m^2)) R(M^{-1}(p,y_2,\dots,y_m)^T)\,dy_2\dots dy_m,
\end{align*}
where this last expression corresponds to the integral on the ball $\mathbb B(0, (1-p^2)^{\frac{1}{2}})$ in $\R^{m-1}$ with respect to the vector $ (y_2,\dots,y_m)$. Applying the Pizzetti formula (\ref{Pi on Ball}) on this ball and following a similar reasoning as in the previous case, we obtain:
\begin{align*}
\int_{\mathbb B_{p,\underline{\omega}}} R(\underline{x}) \, d S_{\underline{x}}&=\sum_{k=0}^\infty\frac{\pi^{\frac{m-1}{2}}}{2^{2k}k!\,\Gamma(\frac{m-1}{2}+1+k)}(\Delta_{\underline{y}}-\partial_{y_1}^2)^k R(M^{-1}(p,y_2,\dots,y_m)^T)\Big|_{y_2,\ldots,y_m=0} (1-p^2)^{k+\frac{m-1}{2}}\nonumber \\
&=\sum_{k=0}^\infty\frac{\pi^{\frac{m-1}{2}}}{2^{2k}k!\,\Gamma(\frac{m+1}{2}+k)}\left(\Delta_{\underline{x}}-\langle\m{\omega}, \pa_{\m{x}}\rangle^2\right)^k [R]\Big|_{\underline{x}=p\underline{\omega}} (1-p^2)^{k+\frac{m-1}{2}},
\end{align*}
which proves the second part of Theorem \ref{PizzMainTheo}, $ii)$.

\subsection{Integration on the Spherical Caps $C_{p,\underline{\omega}}$ and $C'_{p,\underline{\omega}}$}
We now discuss the cases of the complementary spherical caps $C_{p,\underline{\omega}}$ and $C'_{p,\underline{\omega}}$, given by the intersection of $\Sa^{m-1}$ with the upper and lower half of $\R^{m+1}$ with respect to $H_{\m{\omega},p}$ respectively, {see Figure 2}. It is enough to explain the details of the case of $C_{p,\underline{\omega}}$, since integration on $C'_{p,\underline{\omega}}$ can be treated in an analogous way. We first recall that:
\[
C_{p,\underline{\omega}}=\{\underline{x}\in\mathbb R^m: \, \langle \underline{x},\underline{\omega}\rangle >p,\, \| \underline{x}\|=1\}, \;\;\;\;\;\;\;   \m{\omega}\in\Sa^{m-1}, \;\;0\leq p<1.
\]
Clearly $C_{p,\underline{\omega}}$ is a $(m-1)-$dimensional surface on $\mathbb \Sa^{m-1}$ and, by virtue of (\ref{NO_Int_For_Heav}), we obtain the following   integration formula:
\[
\int_{C_{p,\underline{\omega}}} R(\underline{x}) \, d S_{\underline{x}}=2\int_{\mathbb R^m} 
H({\langle \underline{x},\underline{\omega}\rangle}-p)
\delta(\|\underline{x}\|^2-1) R(\underline{x})\,dV_{\underline{x}}\, .
\]
The same type of rotation $\underline{y}=M \underline{x}$ (where $M\in SO(m)$ is such that $y_1=\langle \underline{x},\underline{\omega}\rangle$) yields:
\begin{align*}
\int_{C_{p,\underline{\omega}}} R(\underline{x}) \, d S_{\underline{x}}&=2\int_{\mathbb R^m} H(y_1-p)
\delta(\|\underline{y}\|^2-1) R(M^{-1}\underline{y})\,dV_{\underline{{y}}}\\
&=\int_p^1 \left(  2\int_{\mathbb R^m} 
\delta(y_2^2+\dots+y_m^2-(1-y_1^2)) \, R(M^{-1}(y_1,y_2,\dots, y_m)^T) \,dy_2 \dots dy_m\right) dy_1.
\end{align*}
The expression between the brackets above has been already computed in (\ref{IntInt}). This expression coincides with the integral over the sphere $(1-y_1^2)^{\frac{1}{2}}\mathbb S^{m-2}\subset \mathbb R^{m-1}$ with respect to  $(y_2,\dots, y_m)$, up to the factor $(1-y^2_1)^{-1/2}$ {(see (\ref{DistIntSph}))}. Pizzetti's formula (\ref{Pi on r-Sph}) then yields:
\begin{align}\label{Pi on Cap}
\int_{C_{p,\underline{\omega}}} \hspace{-.3cm}R(\underline{x}) \, d S_{\underline{x}}&=(1-y_1^2)^{-\frac{1}{2}}\int_p^1 \left(\int_{(1-y_1^2)^{\frac{1}{2}}\mathbb S^{m-2}} R(M^{-1}(y_1,y_2,\dots, y_m)^T) dS_{(y_2,\ldots, y_m)}\right) dy_1 \nonumber\\
&= \sum_{k=0}^\infty\frac{2\pi^{\frac{m-1}{2}}}{2^{2k}k!\,\Gamma(\frac{m-1}{2}+k)} \int_p^1 
(\Delta_{\underline{y}}-\partial_{y_1}^2)^k R(M^{-1}(y_1,y_2,\dots,y_m)^T)\Big|_{y_2,\ldots,y_m=0} (1-y_1^2)^{k+\frac{m-3}{2}} dy_1\nonumber\\
&=\sum_{k=0}^\infty\frac{2\pi^{\frac{m-1}{2}}}{2^{2k}k!\,\Gamma(\frac{m-1}{2}+k)}\int_p^1 
\left(\Delta_{\underline{x}}-\langle\m{\omega}, \pa_{\m{x}}\rangle^2\right)^k [R]\Big|_{\underline{x}=y_1\underline{\omega}}(1-y_{{1}}^2)^{k+\frac{m-3}{2}}\,dy_1 .
\end{align}
Formula (\ref{Pi on Cap}) provides a Pizzetti formula for integrating arbitrary polynomials $R\in\mathcal{P}(\R^m)$ over the spherical cap  $C_{p,\underline{\omega}}$. Contrary to the previous Pizzetti-type formulae, the above expression does not depend only on the action of invariant differential operators on the integrand. Instead, it combines this action with a 1-dimensional integral with respect to the component $y_1=\langle\m{x},\m{\omega}\rangle$. This integral can {be further} simplified if we restrict these actions to the homogeneous components of $R$.

Indeed, let us assume that $R=R_\el$ is an homogeneous polynomial of degree $\el\in\N_0$. This assumption involves no loss of generality since we can always decompose polynomials into their homogeneous components, see (\ref{Dec1}). In this case, we have that $\left(\Delta_{\underline{x}}-\langle\m{\omega}, \pa_{\m{x}}\rangle^2\right)^k [R_\el]$ is homogeneous of degree $\el-2k$ for $\ell\ge 2k$. Then one obtains:
\[
\left(\Delta_{\underline{x}}-\langle\m{\omega}, \pa_{\m{x}}\rangle^2\right)^k [R_\el]\Big|_{\underline{x}=y_1\underline{\omega}} 
=
\begin{cases}
y_1^{\el-2k} \left(\Delta_{\underline{x}}-\langle\m{\omega}, \pa_{\m{x}}\rangle^2\right)^k [R_\el]\Big|_{\underline{x}=\underline{\omega}}\, , & \el\ge 2k,\\
0\, , & \el< 2k.
\end{cases}
\]
Therefore, on the space $\mathcal{P}_\el(\R^m)$ of homogenous polynomials of degree $\el$, we can re-write (\ref{Pi on Cap}) as
\[
\int_{C_{p,\underline{\omega}}} R_\el(\underline{x}) \, d S_{\underline{x}}=\sum_{k=0}^{\lfloor \frac{\el}{2}\rfloor}\frac{2\pi^{\frac{m-1}{2}}}{2^{2k}k!\,\Gamma(\frac{m-1}{2}+k)}
c_{k,\el}(p) \left(\Delta_{\underline{x}}-\langle\m{\omega}, \pa_{\m{x}}\rangle^2\right)^k [R_\el]\Big|_{\underline{x}=\underline{\omega}} \, ,
\]
where $c_{k,\el}(p):=\displaystyle \int_p^1 y_1^{\el-2k}(1-y^2)^{k+\frac{m-3}{2}}\,dy_1$. 

The integral coefficients $c_{k,\el}(p)$ in the above summation can be computed explicitly. Indeed, let us first recall that a primitive for the function $f(t)=t^a(1-t^2)^b$ is given by
\[
\frac{t^{a+1}}{a+1} \; {}_2F_1\left(-b,\frac{a+1}{2};\frac{a+3}{2};t^2\right), 
\]
where ${}_2F_1(a,b;c,z)$ is the hypergeometric function in the variable $z$ of parameters $(a,b)$ and $c$.
Combining this fact with the well-known Gauss summation theorem for hypergeometric functions (see Theorem 2.2.2 in \cite{MR1688958}), i.e.\
\begin{equation}\label{GST}
{}_2F_1(a,b;c,1)=\frac{\Gam(c) \Gam(c-b-a)}{\Gam(c-a)\Gam(c-b)}, \;\;\;\; \;\; c>a+b, \;\;\;\;  a,b,c\in\R,
\end{equation}
we obtain that 
\begin{align*}
\int_p^1 t^a(1-t^2)^b \, dt &=  \frac{t^{a+1}}{a+1} \, {}_2F_1\left(-b,\frac{a+1}{2};\frac{a+3}{2};t^2\right) \Bigg|^{t=1}_{t=p} \\
&= \frac{\Gam\left(\frac{a+1}{2}\right) \Gam\left(b+1\right)}{2\Gam\left(\frac{a+3}{2}+b\right)} - \frac{p^{a+1}}{a+1} \, {}_2F_1\left(-b,\frac{a+1}{2};\frac{a+3}{2};p^2\right) \, .
\end{align*}
Finally substituting $a=l-2k$ and $b=k+\frac{m-3}{2}$ we obtain that 
\begin{align*}
c_{k, \el}(p) &:= \int_p^1 y_1^{\el-2k}(1-y_1^2)^{k+\frac{m-3}{2}}\,dy_1 \\
&= \frac{\Gamma\left(\frac{\el+1}{2}-k\right)\Gamma\left(k+ \frac{m-1}{2}\right)}{2\Gamma\left( \frac{\el+m}{2}\right)} - \frac{p^{\el-2k+1}}{\el-2k+1} {}_2F_1\left(-k-\frac{m-3}{2}, \frac{\el+1}{2}-k;\frac{\el+3}{2}-k, p^2\right)\, ,
\end{align*}
which proves  the third part of the Theorem \ref{PizzMainTheo} $iii)$. As mentioned above, similar arguments apply to the proof of the fourth part $iv)$.


\section{Application: Inversion {Formulae} for the Spherical Radon Transform}
\label{Radon Sperical}
In this section we provide alternative proofs to the inversion formulae of the ($m-2$)-dimensional spherical Radon transform (also known as the Funk transform). In particular, we will prove Theorems \ref{Helgason spherical inversion, general k even} and \ref{Helgason2general} for $k=1$ by means of direct computations using the Pizzetti formulae in Theorem \ref{PizzMainTheo}.

The spherical Radon transform in question is an integral transformation that maps functions in $L^2(\mathbb S^{m-1})$ to functions in $L^2(\Xi)$, where $\Xi$ is the manifold of all $(m-2)$-dimensional totally geodesic sub-spheres of $\mathbb S^{m-1}$. Given a function $f\in L^2(\mathbb S^{m-1})$, its spherical Radon transform $\widehat{f}$ is defined as
\begin{align}
\label{ART}
\widehat{f}(\xi)=\int_{\xi} f(\underline{x})d S_{\underline{x}}\, ,\qquad \xi\in  \Xi.
\end{align}
Since the above transform is taken over sub-spheres of co-dimension 1 with respect to $\Sa^{m-1}$, the formula above can be rewritten in simpler terms. Indeed, it is easily seen that for every $\xi\in  \Xi$ there exists $\m{\omega}\in\Sa^{m-1}$ such that ${\m{\omega}}\perp \xi$, i.e.\
\[
\xi= \Sa_{0,\m{\omega}}= \{\m{x}\in\Sa^{m-1}: \langle \m{x}, \m{\omega}\rangle=0\}.
\]
In other words, in the case $k=1$, the map (\ref{St-SubS}) defines a double covering of $\Xi$ by $\Sa^{m-1}$ by means of the identification
\begin{equation}\label{dc}
\m{\omega} \;\; \mapsto \;\;\Sa_{0,\m{\omega}}:= \{\m{x}\in\Sa^{m-1}: \langle \m{x}, \m{\omega}\rangle=0\}.
\end{equation}
It is readily seen that every pair of antipodal unit vectors $\underline{\omega}$ and $-\underline{\omega}$ in $\mathbb S^{m-1}$ define the same $(m-2)$-dimensional geodesic sphere $\Sa^{m-2}_{\perp\m{\omega}}$.
\begin{remark}
In the general case $1\leq k\leq m-2$, the map (\ref{St-SubS}) defines a \textup{O}$(k)$-invariant mapping from St$(m,k)$ onto $\Xi$. To see this, it is enough to consider the natural group action {of \textup{O}$(k)$ on \textup{St}$(m,k)$}:
 \[
\textup{St}(m,k)\times \textup{O}(k)\fd \textup{St}(m,k): M\mapsto Mg, \;\;\;\; M\in \textup{St}(m,k), \; g\in  \textup{O}(k).\]
{Here} we are seeing elements in  St$(m,k)$ as matrices of $k$ orthonormal column vectors in $\R^m$, i.e.  St$(m,k)=\{M\in \R^{m\times k}: M^TM=\I_k\}$. {It} is clear that any element of the orbit $\{Mg: g\in \textup{O}(k)\}$ of $M\in \textup{St}(m,k)$ defines the same $(m-k-1)$-dimensional subshpere via the mapping (\ref{St-SubS}), {which establishes the \textup{O}$(k)$-invariance.}
\end{remark}

In view of the double covering (\ref{dc}), we can see any function $\phi$ in $\Xi$ as an even function defined on $\Sa^{m-1}$. By virtue of this reasoning, and by abuse of notation, we can rewrite formula (\ref{ART}) (see also (\ref{RadvsStief})) as
\begin{align}
\label{RT}
\widehat{f}(\underline{\omega})=\int_{\Sa_{0,\m{\omega}}} f(\underline{x}) \, d S_{\underline{x}}\, ,\qquad \underline{\omega}\in \mathbb S^{m-1}.
\end{align}


Along with the transformation  $f\mapsto \widehat{f}$ we shall consider its dual transform $\phi \mapsto \widecheck{\phi}$, which {to a function  $\phi\in L^2(\Xi)$ associates the function $\widecheck{\phi}\in L^2(\Sa^{m-1})$ given by}
\[
\widecheck{\phi}(\m{x}) = \int_{\xi\ni\m{x}} \phi(\xi) \, d\mu(\xi)\, ,
\]
where $d\mu$ is the nomalized invariant measure on the set $\{\xi\in\Xi: \m{x}\in\xi\}$. In view of the double-covering (\ref{dc}), we can repeat a {reasoning similar to the one used for the Radon transform}.
In particular, it is easily seen that set of $(m-2)-$dimensional geodesic spheres passing through $\underline{x}\in\Sa^{m-1}$ is double covered by the geodesic sub-sphere $\Sa_{0,\m{x}}$ orthogonal to $\m{x}$. Thus the dual transform of $\phi$ can be written as (see (\ref{dualStf})):
\begin{align}
\label{RT phi dual}
\widecheck{\phi}(\underline{x})=\frac{1}{\sigma_{m-1}}\int_{\Sa_{0,\m{x}}} \phi(\underline{\omega}) \, d S_{\underline{\omega}}\, .
\end{align}
\begin{remark}
Comparing formulae  (\ref{RT}) and (\ref{RT phi dual}), it is easily seen that the spherical Radon transform, defined on geodesic sub-spheres of $\Sa^{m-1}$ of codimension one, is (up to a constant factor) identical to its dual transform.
\end{remark}

Before we proceed to the inversion formulae for the spherical Radon transform, we need the following observation. Each geodesic sub-sphere in $\Sa^{m-1}$  through a point $\underline{x}$ also passes through the antipodal point $A\underline{x}=-\underline{x}$. It is thus clear that $\widehat{f}=\widehat{f\circ A}$ and, as a result, $\widehat{f}=0$ if $f$ is an odd function on $\mathbb S^{m-1}$, i.e. if $f+f\circ A=0$. In \cite[Thm 4.7]{MR754767}, it was shown that the kernel of $\widehat{\cdot}$ is exactly the set of odd functions on $\mathbb S^{m-1}$. So, when considering the problem of inverting this transform, it is natural to confine our attention to the case of even functions.

\subsection{Inversion {formula}, case $m$ even}

The inversion formula provided in Theorem \ref{Helgason spherical inversion, general k even} was established by Helgason (see e.g.\ \cite[Thm.~1.17 Ch.~3.1]{MR2743116} and  \cite[Thm 4.7]{MR754767}) under the assumption that $m-k-1$ is even. In the particular case where $k=1$, this result reads as follows:
\begin{teo}
\label{Helgason spherical inversion}
Let $m\in \N$ be even. The spherical Radon transform $f\longrightarrow \widehat{f}$ (for an even function $f$) is inverted by the formula
\begin{align}
\label{Helgason formula}
2(-4\pi)^{\frac{m-2}{2}} \frac{\Gam\left(\frac{m-1}{2}\right)}{\Gam\left(\frac{1}{2}\right)} \, f=P_{m-2}(\Delta_{LB}) \; (\widehat{f}) \widecheck{\phantom{f}},
\end{align}
where $\Delta_{LB}=r^2\Del_{\m{x}}-(m-2+\mathbb E)\mathbb E$ is the Laplace-Beltrami operator on $\mathbb S^{m-1}$ and $P_{m-2}$ is the polynomial
\begin{align*}
P_{m-2}(z)=\prod_{j=1}^{\frac{m-2}{2}} \left[z-(m-2j-1)(2j-1)\right].
\end{align*}
\end{teo}

{We shall now use} Theorem \ref{PizzMainTheo} to {directly verify (\ref{Helgason formula})}.
To that end, we first recall that $L^2(\mathbb S^{m-1})$ admits the following orthogonal  decomposition into spaces of spherical harmonics (see Proposition \ref{OrthDecSH}) }
\begin{align*}
L^2(\mathbb S^{m-1})=\bigoplus_{k=0}^{\infty} \mathcal H_{k}(\Sa^{m-1}).
\end{align*}
In view of this decomposition, it is enough to prove that Theorem \ref{Helgason spherical inversion} holds for elements of $ \mathcal H_{k}(\Sa^{m-1})$, $k\in\N$. {Let} us consider then $H_k\in \mathcal H_k$. Using Pizzetti's formula (\ref{Pi on S}) for $p=0$, we have:
\begin{align}\label{26}
\widehat{H_{2k}}(\underline{\omega})&=\int_{{\mathbb S_{0,\underline{\omega}}}} H_{2k}(\underline{x}) \, d S_{\underline{x}}\nonumber\\
&=\frac{2\,\pi^{\frac{m-1}{2}}}{2^{2k}k!\,\Gamma\left(\frac{m-1}{2}+k\right)}
\left(\Delta_{\underline{x}}-\langle \underline{\omega},\pa_{\underline{x}}\rangle^2\right)^k \,H_{2k}({\underline{x}}) \nonumber \\
&=\frac{2\,\pi^{\frac{m-1}{2}}}{2^{2k}k!\,\Gamma(\frac{m-1}{2}+k)}
(-1)^k \langle \underline{\omega}, \partial_{\underline{x}}\rangle^{2k} \,H_{2k}({\underline{x}})\, .
\end{align}
We now shall make use of the following result:
\begin{lem}\label{PolDDer}
Let $j,\el\in\N_0$ be such that $j\leq \el$, and let $P_\el \in \mathcal{P}_\el(\R^m)$. Then
\begin{equation}\label{HomPro2}
{\langle \underline{\omega},\pa_{\underline{x}}\rangle}^{j} [P_\el]\Big|_{\underline{x}=\underline{\omega}} = \frac{\el!}{(\el-j)!} P_\el(\m{\omega})\, . 
\end{equation}
\end{lem}
\noindent {\em Proof of Lemma \ref{PolDDer}.}  It is enough to prove this result for the basis elements $\m{x}^\be := x_1^{\be_1}\cdots x_m^{\be_m}$ of  $\mathcal{P}_\el(\R^m)$, where $\be=(\be_1,\ldots, \be_m)\in\N_0^m$ is a multi-index such that $|\be|:=\be_1+\cdots+\be_m=\el$. For these elements, we have
\begin{align*}
{\langle \underline{\omega},\pa_{\underline{x}}\rangle}^{j} [\m{x}^\be]\Big|_{\underline{x}=\m{\omega}} &=\sum_{|\alpha |=j}\frac{j!}{\alpha_1 ! \cdots \al_m!} \,  \underline{\omega}^{\alpha} \, \pa_{x_1}^{\al_1}\cdots \pa_{x_m}^{\al_m} [\underline{x}^{\beta}] \Big|_{\underline{x}=\m{\omega}} \, \\
&= \sum_{|\alpha |=j \atop \al\leq \be}\frac{j!}{\alpha_1 ! \cdots \al_m!} \,  \underline{\omega}^{\alpha} \, \frac{\be_1!}{(\be_1-\al_1)!} \cdots \frac{\be_m!}{(\be_m-\al_m)!} \, \omega_1^{\be_1-\al_1} \cdots \omega_m^{\be_m-\al_m}  \\
&=j!  \binom{\be_1+\cdots +\be_m}{\al_1 +\cdots +\al_m}\m{\omega}^\be \\
&= \frac{\el!}{(\el-j)!} \, \m{\omega}^\be.
\end{align*}
$\hfill\square$

Lemma \ref{PolDDer} clearly yields the {identity} $\langle \underline{\omega}, \partial_{\underline{x}}\rangle^{2k} \,H_{2k}({\underline{x}}) = (2k)! \, H_{2k}({\underline{w}})$. Substituting this into the formula (\ref{26}) {we obtain}
\begin{equation}\label{RadTraHar}
\widehat{H_{2k}}(\underline{\omega})=2\,\pi^{\frac{m-1}{2}} \, \frac{ (-1)^k(2k!)}{2^{2k}k!\,\Gamma\left(\frac{m-1}{2}+k\right)}
\,H_{2k}({\underline{\omega}}) = 2\pi^{\frac{m}{2}-1} \frac{(-1)^k \Gamma\left(k+\frac{1}{2}\right)}{\Gamma\left(\frac{m-1}{2}+k\right)}\,H_{2k}({\underline{\omega}})\, ,
\end{equation}
where  in the last equality we have used the identity $\displaystyle \frac{(2k)!}{2^{2k}\, k!}=  \frac{\Gamma(k+\frac{1}{2})}{\pi^{\frac{1}{2}}}$. Repeating the same reasoning with the dual transform $\phi\mapsto \widecheck{\phi}$, we obtain:
\begin{align}
\label{H RT Dual}
{(\widehat{H_{2k}})}^\vee(\underline{x})
&=\frac{4\pi^{m-2}}{\sigma_{m-1}} \, \frac{\Gamma(k+\frac{1}{2})^2}{\Gamma\left(\frac{m-1}{2}+k\right)^2} \,H_{2k}({\underline{x}})\, .
\end{align}

\noindent We now recall that $H_{2k}$ is an eigenvector of the Laplace-Beltrami operator with eigenvalue $-2k(m-2+2k)$. Therefore the action of $P_{m-2}(\Delta_{LB})$ on both sides of (\ref{H RT Dual}) yields:
\begin{align}
\label{P LB}
P_{m-2}(\Delta_{LB})\, {(\widehat{H_{2k}})}^{\vee}
&=\frac{4\pi^{m-2}}{\sigma_{m-1}} \frac{\Gamma(k+\frac{1}{2})^2}{\Gamma(\frac{m-1}{2}+k)^2} \, C_{k,m} \, H_{2k}\, ,
\end{align}
where the constant $C_{k,m}$ is determined by the action of  $P_{m-2}(\Delta_{LB})$ on $H_{2k}$, or equivalently, 
\[\displaystyle C_{k,m}=P_{m-2}\left(-2k(m-2+2k)\right) = \prod_{j=1}^{\frac{m-1}{2}} [-2k(m-2+2k)-(m-2j-1)(2j-1)]\, .\]
To explicitly compute $C_{k,m}$ we first note that
\begin{align}\label{29}
P_{m-2}(z)&=\prod_{j=1}^{\frac{m-2}{2}} \big[z-(m-2j-1)(2j-1)\big] \nonumber\\
&=4^{\frac{m}{2}-1} \prod_{j=1}^{\frac{m}{2}-1} \left(j^2-\frac{m}{2} j+\frac{z+m-1}{4}\right)\nonumber\\
&=4^{\frac{m}{2}-1} \prod_{j=1}^{\frac{m}{2}-1} (j-j_+)(j-j_-) \nonumber\\
&=4^{\frac{m}{2}-1}\, \frac{\Gamma\left(\frac{m}{2}-j_+\right)}{\Gamma(1-j_+)}\frac{\Gamma\left(\frac{m}{2}-j_-\right)}{\Gamma(1-j_-)},
\end{align}
where
\[
j_{\pm}=\frac{\frac{m}{2}\pm\sqrt{\frac{m^2}{4}-z+m-1}}{2}=\frac{1}{4}\left(m \pm \sqrt{m^2-4(z+m-1)} \,\right)
\]
are the roots of the quadratic polynomial $\displaystyle j^2-\frac{m}{2} j+\frac{z+m-1}{4}$ in the variable $j$. 

\noindent In the case where $z=-2k(m-2+2k)$, we have that $m^2-4(z+m-1)=(m+4k-2)^2$. Thus the roots $j_{\pm}$ can be computed in this case as
\[
j_+=\frac{m}{2}+k-\frac{1}{2}, \;\;\;\; \mbox{ and } \;\;\;\; j_-=-k+\frac{1}{2},
\]
respectively. Therefore
\begin{align*}
C_{k,m}=4^{\frac{m}{2}-1} \frac{\Gamma(\frac{1}{2}-k)}{\Gamma(\frac{3-m}{2}-k)}\frac{\Gamma(\frac{m-1}{2}+k)}{\Gamma(\frac{1}{2}+k)},
\end{align*}
which {yields the following equality when substituted in (\ref{P LB})},
\begin{align}\label{28}
P_{m-2}(\Delta_{LB})\, {(\widehat{H_{2k}})}^{\vee}
&=\frac{4^{\frac{m}{2}}\pi^{m-2}}{\sigma_{m-1}} \, \frac{\Gamma(k+\frac{1}{2})\Gamma(\frac{1}{2}-k)}{\Gamma(\frac{3}{2}-\frac{m}{2}-k) \Gamma(\frac{m}{2}+k-\frac{1}{2})} \, H_{2k}.
\end{align}
Using the identities $\displaystyle\Gamma\left(z+\frac{1}{2}\right)\Gamma\left(\frac{1}{2}-z\right)=\frac{\pi}{\cos(\pi z)}$ and $\displaystyle\Gamma(z)\Gamma(-z+1)=\frac{\pi}{\sin(\pi z)}$, we easily obtain that
\[
\Gamma\left(k+\frac{1}{2}\right)\Gamma\left(\frac{1}{2}-k\right)=(-1)^k\,\pi \;\;\; \mbox{ and } \;\;\; \Gamma\left(\frac{3}{2}-\frac{m}{2}-k\right) \Gamma\left(\frac{m}{2}+k-\frac{1}{2}\right)=(-1)^{\frac{m}{2}+k-1}\,\pi,
\]
respectively. Finally, substituting these identities into (\ref{28}) yields
\begin{align*}
P_{m-2}(\Delta_{LB})(\widehat{H_{2k}})^{\vee}=(-1)^{\frac{m}{2}-1} \, \frac{4^{\frac{m}{2}}\pi^{m-2}}{\sigma_{m-1}}  \, H_{2k} = 2(-4\pi)^{\frac{m-2}{2}} \frac{\Gam\left(\frac{m-1}{2}\right)}{\Gam\left(\frac{1}{2}\right)} \, H_{2k},
\end{align*}
which {completes the proof of Theorem \ref{Helgason spherical inversion}.} 


\subsection{Inversion {formula}, the general case}
\label{SubInvGenk1}

In this section, we further apply Pizzetti formulae to the problem of inverting the spherical Radon transform. In particular, we will prove Theorem \ref{Helgason2general} for the case $k=1$, which constitutes an extension of Theorem \ref{Helgason spherical inversion} to any arbitrary dimension $m\in\N$.


To that end, we first introduce the following generalization of the dual Radon transform (\ref{RT phi dual}).  Given $\phi\in L^2(\Xi)$ and $0\leq r\leq\frac{\pi}{2}$, we denote by $\widecheck{\phi}_r(\m{x})$ the {\em average} of $\phi$ over all $(m-2)-$dimensional geodesic sub-spheres passing at  a distance $r$ from $\underline{x}\in\Sa^{m-1}$ i.e.\
\begin{align}
\label{avg}
\widecheck{\phi}_r(\underline{x})=\int_{d(\underline{x},\xi)=r} \phi(\xi) \, d\mu(\xi)\qquad \qquad  \phi\in L^2(\Xi),
\end{align}
where $d(\underline{x},\xi)$ is the geodesic distance between $\m{x}$ and the sub-sphere $\xi$ and  $d\mu$ is the normalized invariant measure on the set $\{\xi\in \Xi \, :\, d(\underline{x},\xi)=r\}$.  Observe that taking $r=0$ yields the dual transform 
$\widecheck{\phi}$ defined in (\ref{RT phi dual}). We {can} now state the following version of Theorem \ref{Helgason2general} for the particular case $k=1$.
\begin{teo}
\label{Helgason2}
The Spherical Radon transform $f\mapsto \widehat{f}$ is, for even functions $f$,   inverted by
\begin{align}
\label{k=m-2}
f(\underline{x})=\frac{2^{m-2}}{(m-3)!\, \sigma_{m-1}} \, \left(\frac{d}{dt^2}\right)^{m-2} \left[ \int_0^t {(\widehat{f})}^{\vee}_{\cos^{-1}(q)}(\underline{x}) \, q^{m-2} \, (t^2-q^2)^{\frac{m-4}{2}}\, dq \right]_{t=1}.
\end{align}
\end{teo}

In order to prove this result using Pizzetti formulae, we need the following crucial observation. In view of the double covering (\ref{dc}) of $\Xi$ by $\Sa^{m-1}$, formula (\ref{avg})  can be rewritten {in terms of} a spherical integral. Indeed, let $\underline{y}\in\mathbb{S}^{m-1}$ be such that $d(\underline{y},\underline{x})=r$, and let $\xi\in \Xi$ be the sub-sphere passing through $\underline{y}$ such that $d(\xi,\underline{x})=r$, see {Figure 3}. It is easily seen that one of the two normal vectors defining $\xi$ satisfies the condition 
\begin{equation}\label{dist}
d(\underline{\omega}, \underline{x})=\frac{\pi}{2}-r.
\end{equation}

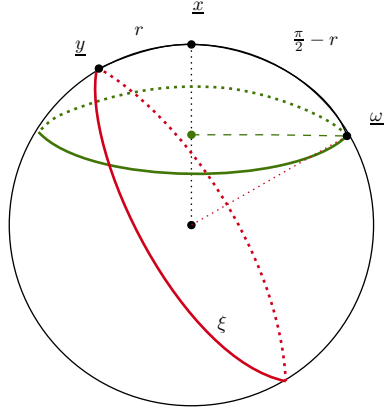
\begin{figure}[H]
       \centering
        \resizebox{0.33\textwidth}{!}{
        \tikzset{every picture/.style={line width=0.75pt}} 

\begin{tikzpicture}[x=0.75pt,y=0.75pt,yscale=-1,xscale=1]

\draw  [color={rgb, 255:red, 0; green, 0; blue, 0 }  ,draw opacity=1 ][line width=0.75]  (3,174) .. controls (3,98.34) and (64.34,37) .. (140,37) .. controls (215.66,37) and (277,98.34) .. (277,174) .. controls (277,249.66) and (215.66,311) .. (140,311) .. controls (64.34,311) and (3,249.66) .. (3,174) -- cycle ;
\draw [color={rgb, 255:red, 0; green, 0; blue, 0 }  ,draw opacity=1 ][line width=0.75]  [dash pattern={on 0.84pt off 2.51pt}]  (140,37) -- (140,174) ;
\draw [shift={(140,174)}, rotate = 90] [color={rgb, 255:red, 0; green, 0; blue, 0 }  ,draw opacity=1 ][fill={rgb, 255:red, 0; green, 0; blue, 0 }  ,fill opacity=1 ][line width=0.75]      (0, 0) circle [x radius= 2.68, y radius= 2.68]   ;
\draw [shift={(140,37)}, rotate = 90] [color={rgb, 255:red, 0; green, 0; blue, 0 }  ,draw opacity=1 ][fill={rgb, 255:red, 0; green, 0; blue, 0 }  ,fill opacity=1 ][line width=0.75]      (0, 0) circle [x radius= 2.68, y radius= 2.68]   ;
\draw [color={rgb, 255:red, 65; green, 117; blue, 5 }  ,draw opacity=1 ][line width=1.5]    (257,106.4) .. controls (216,146) and (62,144.4) .. (25,103.4) ;
\draw [color={rgb, 255:red, 65; green, 117; blue, 5 }  ,draw opacity=1 ][line width=1.5]  [dash pattern={on 1.69pt off 2.76pt}]  (257,106.4) .. controls (248.5,94.94) and (218.79,82.31) .. (196.51,76.17) .. controls (174.23,70.03) and (160.76,68.96) .. (141,68.96) .. controls (121.24,68.95) and (94.21,71.76) .. (64.45,81.73) .. controls (34.68,91.7) and (34.85,93.89) .. (25,103.4) ;
\draw [color={rgb, 255:red, 65; green, 117; blue, 5 }  ,draw opacity=1 ][line width=0.75]  [dash pattern={on 4.5pt off 4.5pt}]  (140,105.5) -- (198,105.5) -- (257,106.4) ;
\draw [shift={(140,105.5)}, rotate = 0] [color={rgb, 255:red, 65; green, 117; blue, 5 }  ,draw opacity=1 ][fill={rgb, 255:red, 65; green, 117; blue, 5 }  ,fill opacity=1 ][line width=0.75]      (0, 0) circle [x radius= 2.68, y radius= 2.68]   ;
\draw [color={rgb, 255:red, 208; green, 2; blue, 27 }  ,draw opacity=1 ][line width=1.5]    (69,54.4) .. controls (56,120.4) and (134,276.4) .. (211,292.4) ;
\draw [color={rgb, 255:red, 208; green, 2; blue, 27 }  ,draw opacity=1 ][line width=1.5]  [dash pattern={on 1.69pt off 2.76pt}]  (69,54.4) .. controls (130,88.4) and (206,193.4) .. (211,292.4) ;
\draw [color={rgb, 255:red, 208; green, 2; blue, 27 }  ,draw opacity=1 ][line width=0.75]  [dash pattern={on 0.84pt off 2.51pt}]  (257,106.4) -- (140,174) ;
\draw [color={rgb, 255:red, 0; green, 0; blue, 0 }  ,draw opacity=1 ][line width=0.75]    (70.4,55.28) .. controls (93.6,44.08) and (120,36.08) .. (140,37) ;
\draw [shift={(70.4,55.28)}, rotate = 334.23] [color={rgb, 255:red, 0; green, 0; blue, 0 }  ,draw opacity=1 ][fill={rgb, 255:red, 0; green, 0; blue, 0 }  ,fill opacity=1 ][line width=0.75]      (0, 0) circle [x radius= 2.68, y radius= 2.68]   ;
\draw [color={rgb, 255:red, 0; green, 0; blue, 0 }  ,draw opacity=1 ][line width=0.75]    (140,37) .. controls (194,37.4) and (246,72.4) .. (257,106.4) ;
\draw [shift={(257,106.4)}, rotate = 72.07] [color={rgb, 255:red, 0; green, 0; blue, 0 }  ,draw opacity=1 ][fill={rgb, 255:red, 0; green, 0; blue, 0 }  ,fill opacity=1 ][line width=0.75]      (0, 0) circle [x radius= 2.68, y radius= 2.68]   ;

\draw (139.6,3) node [anchor=north west][inner sep=0.75pt]    {$\underline{x}$};
\draw (273,87.4) node [anchor=north west][inner sep=0.75pt]    {$\underline{\omega }$};
\draw (51,32.4) node [anchor=north west][inner sep=0.75pt]    {$\underline{y}$};
\draw (95.8,22.4) node [anchor=north west][inner sep=0.75pt]    {$r$};
\draw (214.2,25.2) node [anchor=north west][inner sep=0.75pt]    {$\frac{\pi }{2} -r$};
\draw (158,241.4) node [anchor=north west][inner sep=0.75pt]    {$\xi $};

\end{tikzpicture}  
       }
       \label{Fig3}
      \caption{Depiction of the subspheres $\xi$ and $\mathbb{S}_{\sin(r),\underline{x}}$ in red and green respectively.}
\end{figure}    


Conversely, any unit vector $\underline{\omega}$ satisfying (\ref{dist}) defines, by virtue of the double covering (\ref{dc}), a geodesic sub-sphere $\xi$ such that $d(\xi,\underline{x})=r$. Then the set $\displaystyle\{\xi\in \Xi : d(\underline{x},\xi)=r\}$ is determined by the set  $\left\{\underline{\omega}\in\mathbb{S}^{m-1} : d(\underline{\omega}, \underline{x})=\frac{\pi}{2}-r\right\}$, which in turn coincides with  the sub-sphere
\begin{align*}
\mathbb{S}_{\sin(r),\underline{x}}\,=\,\left\{\underline{\omega}\in\mathbb{S}^{m-1} : \langle\underline{x},\underline{\omega}\rangle=\sin(r)\right\},
\end{align*}
defined in Theorem \ref{PizzMainTheo} $i)$. This geometric argument thus yields
\begin{align}
\label{phivee}
\widecheck{\phi}_r(\underline{x}) =\frac{1}{\cos^{m-2} (r)\, \sigma_{m-1}}\, \int_{\mathbb{S}_{\sin(r),\underline{x}}} \, \phi(\underline{\omega})\, dS_{\underline{\omega}}.
\end{align}

\noindent {\it Proof of Theorem  \ref{Helgason2}:}

\noindent Due to the decomposition of $L^2(\mathbb{S}^{m-1})$-functions into spherical harmonics, it is enough to prove the theorem for $f=H_{2k} \in \mathcal{H}_{2k}(\Sa^{m-1})$. From (\ref{RadTraHar}) we know that 
\[
\widehat{H_{2k}}(\underline{\omega}) = d_{m,k}\,H_{2k}({\underline{\omega}}),  \;\;\; \mbox{ with }  \;\;\; d_{m,k}= 2\pi^{\frac{m}{2}-1} \frac{(-1)^k \Gamma\left(k+\frac{1}{2}\right)}{\Gamma\left(\frac{m-1}{2}+k\right)}.
\]
Hence, from (\ref{phivee}) we obtain
\[
{(\widehat{H_{2k}})}^{\vee}_r(\m{x}) =\frac{d_{m,k}}{\cos^{m-2} (r)\, \sigma_{m-1}}\, \, \int_{\mathbb{S}_{\sin(r),\underline{x}}} \, H_{2k}(\underline{\omega})\, dS_{\underline{\omega}}.
\]
Taking $p=\sin(r)$ and $q=\cos(r)$, i.e. $p^2+q^2=1$, by virtue of the Pizzetti formula (\ref{Pi on S}), we obtain 
\begin{align}\label{38}
{\left(\widehat{H_{2k}}\right)}^{\vee}_{\cos^{-1}(q)} (\m{x})&=\frac{d_{m,k}}{q^{m-2} \, \sigma_{m-1}}\, \int_{\mathbb{S}_{p,\underline{x}}} \, H_{2k}(\underline{\omega})\, dS_{\underline{\omega}} \nonumber \\
&=\frac{d_{m,k}}{\sigma_{m-1}}\,\frac{1}{q^{m-2}}\, \sum_{j=0}^k\,  \frac{2\pi^{\frac{m-1}{2}}}{2^{2j}\, j! \,\Gamma\left(\frac{m-1}{2}+j\right)}\,\left(\Delta_{\underline{\omega}}-{\langle \underline{x},\pa_{\underline{\omega}}\rangle}^2\right)^j [H_{2k}] \Big|_{\underline{\omega}=p\underline{x}} \, q^{2j+m-2} \nonumber  \\
&= \frac{d_{m,k}}{\sigma_{m-1}}\, \, \sum_{j=0}^k\,  \frac{2\pi^{\frac{m-1}{2}}(-1)^j}{2^{2j}\, j! \,\Gamma\left(\frac{m-1}{2}+j\right)}\,  {\langle \underline{x},\pa_{\underline{\omega}}\rangle}^{2j}[H_{2k}] \Big|_{\underline{\omega}=p\underline{x}} \, q^{2j}\, .
\end{align}
We now recall that ${\langle \underline{x},\pa_{\underline{\omega}}\rangle}^{2j}[H_{2k}] $ is a homogeneous polynomial of degree $2k-2j$ in the variable $\m{\omega}$,  therefore:
\[
{\langle \underline{x},\pa_{\underline{\omega}}\rangle}^{2j}[H_{2k}] \Big|_{\underline{\omega}=p\underline{x}} = p^{2k-2j} {\langle \underline{x},\pa_{\underline{\omega}}\rangle}^{2j}[H_{2k}] \Big|_{\underline{\omega}=\underline{x}}\, ,
\]
and, applying Lemma \ref{PolDDer}, we obtain:
\[
{\langle \underline{x},\pa_{\underline{\omega}}\rangle}^{2j}[H_{2k}] \Big|_{\underline{\omega}=p\underline{x}} =   \frac{(2k)!}{(2k-2j)!}\,  p^{2k-2j} \, H_{2k}(\underline{x})\, .
\]
Substituting the equality above into (\ref{38}) yields:
\begin{align}\label{compInt}
{\left(\widehat{H_{2k}}\right)}^{\vee}_{\cos^{-1}(q)} (\m{x})&=\frac{d_{m,k}}{\sigma_{m-1}}\, \sum_{j=0}^k\,  \frac{2\pi^{\frac{m-1}{2}}(-1)^j}{2^{2j}\, j! \,\Gamma(\frac{m-1}{2}+j)}\,\frac{(2k)!}{(2k-2j)!}\,  p^{2k-2j} \, H_{2k}(\underline{x}) \, q^{2j} \nonumber\\
&=h_{m,k} \, H_{2k}(\underline{x})\, \sum_{j=0}^k\,  \frac{(-1)^j}{2^{2j}\, j! \,\Gamma(\frac{m-1}{2}+j)}\,\frac{p^{2k-2j} q^{2j}}{(2k-2j)!}\, ,
\end{align}
where we have introduced the new constant
\[
h_{m,k} = 2\pi^{\frac{m-1}{2}}\, (2k)! \, \frac{d_{m,k}}{\sigma_{m-1}} =2\pi^{\frac{m}{2}-1}\,(-1)^k\,(2k)! \, \frac{\Gamma(k+\frac{1}{2})\Gamma(\frac{m-1}{2})}{\Gamma(\frac{m-1}{2}+k)}.
\]
The task is now to compute:
\[
F(\underline{x}):= \left(\frac{d}{dt^2}\right)^{m-2} \left[ \int_0^t {\left(\widehat{H_{2k}}\right)}^{\vee}_{\cos^{-1}(q)} (\m{x}) \,\, q^{m-2} \, (t^2-q^2)^{\frac{m-4}{2}}\, dq \right]_{t=1}.
\]
From (\ref{compInt}) we obtain
\begin{align}
\label{genF}
F(\underline{x})=h_{m,k} \, H_{2k}(\underline{x})\, \sum_{j=0}^k\,  \frac{(-1)^j}{2^{2j}\, j! \,\Gamma(\frac{m-1}{2}+j)(2k-2j)!}\,\left(\frac{d}{dt^2}\right)^{m-2} \left[ I_{k,j}(t)\right] \Big|_{t=1},
\end{align}
where
\[
 I_{k,j}(t):= \int_0^t \left(1-q^2\right)^{k-j} \, q^{2j+m-2} \,\left(t^2-q^2\right)^{\frac{m-4}{2}} \,dq.
\]
Applying the change of coordinates $q=t\sin(\theta)$, we obtain $dq=t\cos(\theta)d\theta$, and
\begin{align*}
I_{k,j}(t)
&= t^{2j+2m-5}\int_0^{\frac{\pi}{2}} \left(1-t^2 \sin^2(\theta)\right)^{k-j} \,\sin^{2j+m-2}(\theta)\, \cos^{m-3}(\theta)\,d\theta.
\end{align*}
Expanding $\left(1-t^2 \sin^2(\theta)\right)^{k-j}$ we now get
\begin{align*}
I_{k,j}(t)
&=\sum_{\el=0}^{k-j} {k-j \choose \el} (-1)^\el\, t^{2j+2\el+2m-5}\,\int_0^{\frac{\pi}{2}} \,\sin^{2j+2\el+m-2}(\theta)\, \cos^{m-3}(\theta)\,d\theta.
\end{align*}
Using the known identity 
\[
\int_0^{\frac{\pi}{2}} \,\sin^{a}(\theta)\, \cos^{b}(\theta)\,d\theta=\frac{\Gamma\left(\frac{a+1}{2}\right)\Gamma\left(\frac{b+1}{2}\right)}{2\,\Gamma\left(\frac{a+b}{2}+1\right)},
\]
for the values $a=2j+2\el+m-2$ and $b=m-3$, we have:
\[
I_{k,j}(t)=\sum_{\el=0}^{k-j} {k-j \choose \el} (-1)^\el\, t^{2j+2\el+2m-5}\,\frac{\Gamma\left(j+\el+\frac{m-1}{2}\right)\Gamma\left(\frac{m-2}{2}\right)}{2\,\Gamma\left(j+\el+m-\frac{3}{2}\right)}.
\]
Moreover, from the identity $\left(\frac{d}{dt^2}\right)^{n}t^{\alpha} = \frac{\Gamma\left(\frac{\alpha}{2}+1\right)}{\Gamma\left(\frac{\alpha}{2}-n+1\right)}\, t^{\alpha-2n}$ we immediately obtain:
\[
\left(\frac{d}{dt^2}\right)^{m-2}t^{2j+2l+2m-5} \Bigg|_{t=1}=\frac{\Gamma\left(j+\el+m-\frac{3}{2}\right)}{\Gamma\left(j+\el+\frac{1}{2}\right)},
\]
which yields:
\begin{align}\label{DerI}
\left(\frac{d}{dt^2}\right)^{m-2} \left[ I_{k,j}(t)\right] \Big|_{t=1}
&=\frac{\Gamma\left(\frac{m-2}{2}\right)}{2}\,\sum_{\el=0}^{k-j} {k-j \choose \el} (-1)^\el\, \,\frac{\Gamma(j+\el+\frac{m-1}{2})}{\Gamma(j+\el+\frac{1}{2})}.
\end{align}
From the definition of the hypergeometric function ${}_2F_1(a,b;c,z)$, we now recall that
\begin{equation*}
{}_2F_1(-n,b;c,z) = \frac{\Gam(c)}{\Gam(b)} \sum_{\el=0}^n (-1)^\el  {n \choose \el} \frac{\Gam(b+\el)}{\Gam(c+ \el)} z^\el, \;\;\;\;\;\;\;\;\;\; n\in\N.
\end{equation*}
{Replacing} $n=k-j$, $b=j+\frac{m-1}{2}$, $c=j+\frac{1}{2}$, $z=1$, and {comparing} the resulting expression {with} (\ref{DerI}), yields:
\[
\left(\frac{d}{dt^2}\right)^{m-2} \left[ I_{k,j}(t)\right] \Big|_{t=1} = \frac{\Gamma\left(\frac{m-2}{2}\right)}{2}\, \frac{\Gam\left(j+\frac{m-1}{2}\right)}{\Gam\left(j+\frac{1}{2}\right)} \; {}_2F_1\left(j-k,j+\frac{m-1}{2};j+\frac{1}{2},1\right).
\] 
Applying the Gauss summation formula (\ref{GST}) we now obtain:
\[
\left(\frac{d}{dt^2}\right)^{m-2} \left[ I_{k,j}(t)\right] \Big|_{t=1} = \frac{\Gamma\left(\frac{m-2}{2}\right)}{2}\, \frac{\Gam\left(j+\frac{m-1}{2}\right)}{\Gam\left(k+\frac{1}{2}\right)} \, \frac{\Gam\left(k-j+\frac{2-m}{2}\right)}{\Gam\left(\frac{2-m}{2}\right)}. \] 
Substituting this result back into (\ref{genF}) we have:

\begin{align}\label{Fseg}
F(\underline{x})
&=\frac{h_{m,k}}{2\,\Gamma\left(k+\frac{1}{2}\right)} \, \frac{\Gamma\left(\frac{m-2}{2}\right)}{\Gamma\left(\frac{2-m}{2}\right)}\, H_{2k}(\underline{x})\, \sum_{j=0}^k\,  \frac{(-1)^j}{2^{2j}\, j!} \,\frac{\Gamma\left(k-j+\frac{2-m}{2}\right)}{(2k-2j)!}.
\end{align}

Applying a similar reasoning as before, we can identify the above summation with a hypergeometric function. Indeed, it is easily seen that
\[
\sum_{j=0}^k\,  \frac{(-1)^j}{2^{2j}\, j!} \,\frac{\Gamma\left(k-j+\frac{2-m}{2}\right)}{(2k-2j)!} = \frac{\Gam\left(k+\frac{2-m}{2}\right)}{(2k)!} {}_2F_1\left(-k,-k+\frac{1}{2}; \frac{m}{2}-k,1\right),
\]
which, after applying the Gauss summation formula (\ref{GST}), yields:
\[
\sum_{j=0}^k\,  \frac{(-1)^j}{2^{2j}\, j!} \,\frac{\Gamma\left(k-j+\frac{2-m}{2}\right)}{(2k-2j)!} = \frac{\Gam\left(k+\frac{2-m}{2}\right)}{(2k)!} \frac{\Gam\left(\frac{m}{2}-k\right)\Gam\left(k+\frac{m-1}{2}\right)}{\Gam\left(\frac{m}{2}\right)\Gam\left(\frac{m-1}{2}\right)}.
\]
Using the identity $\frac{\Gam\left(\frac{m}{2}-k\right)}{\Gam\left(\frac{m}{2}\right)}= (-1)^k \frac{\Gam\left(\frac{2-m}{2}\right)}{\Gam\left(k+\frac{2-m}{2}\right)}$ in the previous formula we obtain:

\begin{align*}
\sum_{j=0}^k\,  \frac{(-1)^j}{2^{2j}\, j!} \,\frac{\Gamma(k-j+\frac{2-m}{2})}{(2k-2j)!}=(-1)^k\,\frac{\Gamma(k+\frac{m-1}{2})}{(2k)!}\,\frac{\Gamma(\frac{2-m}{2})}{\Gamma(\frac{m-1}{2})},
\end{align*}
{which in turn allows us to rewrite (\ref{Fseg}) as
}
\begin{align}
F(\underline{x})
&=\frac{h_{m,k}}{2\,\Gamma(k+\frac{1}{2})} \, \frac{\Gamma(\frac{m-2}{2})}{\Gamma(\frac{m-1}{2})}\, (-1)^k\,\frac{\Gamma(k+\frac{m-1}{2})}{(2k)!}\, H_{2k}(\underline{x})\nonumber\\
&=\pi^{\frac{m}{2}-1}\,\Gamma\left(\frac{m-2}{2}\right)\, H_{2k}(\underline{x})\, . \label{Fin}
\end{align}
Finally, applying the Legendre duplication formula for the Gamma function we have that
\[
\Gamma\left(\frac{m-2}{2}\right)\,\Gamma\left(\frac{m-1}{2}\right)=\frac{\pi^{\frac{1}{2}}}{2^{m-3}}\,(m-3)!\, .
\]
Hence, the formula (\ref{Fin}) can be rewritten as 
\begin{align*}
F(\underline{x})
=\frac{2\pi^{\frac{m-1}{2}}}{\Gamma(\frac{m-1}{2})}\,\frac{(m-3)!}{2^{m-2}} \, H_{2k}(\underline{x}) =\frac{\sigma_{m-1} \, (m-3)!}{2^{m-2}}\, H_{2k}(\underline{x}),
\end{align*}
which completes the proof of Theorem \ref{Helgason2}. $\hfill\square$


\appendix\label{Appendix}
\section{Recovering classical Pizzetti formula from integration on spherical caps}\label{App}
In this appendix section, our goal is to provide a direct proof of formula (\ref{Ann}). To that end, let us denote the right hand side of (\ref{Ann}) by
\begin{align*}
S &:= \frac{2^{2s} s!}{\pi^{\frac{1}{2}}} \sum_{k=0}^s \frac{ \Gamma\left(s-k + \frac{1}{2}\right)}{2^{2k}k!}
\left(\Delta_{\underline{x}}-{\langle \underline{\omega},\pa_{\underline{x}}\rangle}^2\right)^k[R_{2s}] \Big|_{\underline{x}=\underline{\omega}} \\
&=  \frac{2^{2s} s!}{\pi^{\frac{1}{2}}} \sum_{k=0}^s \sum_{\el=0}^k \frac{ \Gamma\left(s-k + \frac{1}{2}\right)}{2^{2k}k!} \binom{k}{\el} (-1)^{\el} {\langle \underline{\omega},\pa_{\underline{x}}\rangle}^{2\el} \Delta_{\underline{x}}^{k-\el}  [R_{2s}] \Big|_{\underline{x}=\underline{\omega}}.
\end{align*}
Applying Lemma \ref{PolDDer} to the polynomial  $\Delta_{\underline{x}}^{k-\el}  [R_{2s}] \in \mathcal{P}_{2s+2\el-2k}(\R^m)$ we obtain
\[
{\langle \underline{\omega},\pa_{\underline{x}}\rangle}^{2\el} \Delta_{\underline{x}}^{k-\el}  [R_{2s}] \Big|_{\underline{x}=\underline{\omega}} = \frac{(2s+2\el-2k)!}{(2s-2k)!} \Delta_{\underline{x}}^{k-\el}  [R_{2s}] \Big|_{\underline{x}=\underline{\omega}}.
\]
Hence
\[
S=   \frac{2^{2s} s!}{\pi^{\frac{1}{2}}} \sum_{k=0}^s \sum_{\el=0}^k   \frac{ (-1)^\el \Gamma\left(s-k + \frac{1}{2}\right) (2s+2\el-2k)!}{2^{2k} \, \el!\, (k-\el)!\, (2s-2k)!} \Delta_{\underline{x}}^{k-\el}  [R_{2s}] \Big|_{\underline{x}=\underline{\omega}}.
\]
Taking now $j=k-\el$ and changing the order of summation yields
\begin{align}
S&=   \frac{2^{2s} s!}{\pi^{\frac{1}{2}}} \sum_{k=0}^s \sum_{j=0}^k   \frac{ (-1)^{k-j} \, \Gamma\left(s-k + \frac{1}{2}\right) (2s-2j)!}{2^{2k} \, (k-j)!\, j!\, (2s-2k)!} \Delta_{\underline{x}}^{j}  [R_{2s}] \Big|_{\underline{x}=\underline{\omega}} \nonumber \\
&=\frac{2^{2s} s!}{\pi^{\frac{1}{2}}}  \sum_{j=0}^s \sum_{k=0}^{s-j}  \frac{ (-1)^{k} \, \Gamma\left(s-k -j + \frac{1}{2}\right) (2s-2j)!}{2^{2k+2j} \, k!\, j!\, (2s-2k-2j)!} \Delta_{\underline{x}}^{j}  [R_{2s}] \Big|_{\underline{x}=\underline{\omega}} \nonumber \\
&=\frac{2^{2s} s!}{\pi^{\frac{1}{2}}}  \sum_{j=0}^s \frac{(2s-2j)!}{2^{2j} j!}  \left(\sum_{k=0}^{s-j}  \frac{ (-1)^{k} \, \Gamma\left(s-k -j + \frac{1}{2}\right)}{2^{2k} \, k!\, (2s-2k-2j)!} \right) \Delta_{\underline{x}}^{j}  [R_{2s}] \Big|_{\underline{x}=\underline{\omega}} . \label{AL}
\end{align}
Using the identity $\frac{ \Gamma\left(n + \frac{1}{2}\right)}{(2n)!} = \frac{\pi^{\frac{1}{2}}}{2^{2n} n!}$ for $n=s-k -j$, and by virtue of the binomial theorem, we obtain 
\begin{align*}
\sum_{k=0}^{s-j}  \frac{ (-1)^{k} \, \Gamma\left(s-k -j + \frac{1}{2}\right)}{2^{2k} \, k!\, (2s-2k-2j)!} = \frac{\pi^{\frac{1}{2}}}{2^{2s-2j}} \sum_{k=0}^{s-j} \frac{(-1)^k}{k! (s-j-k)!} = \frac{\pi^{\frac{1}{2}} }{2^{2s-2j} (s-j)!} (1-1)^{s-j}.
\end{align*}
Thus the only non-vanishing term in the sum (\ref{AL}) is the term corresponding to $j=s$. Therefore,
\[
S= \Delta_{\underline{x}}^{s}  [R_{2s}] \Big|_{\underline{x}=\underline{\omega}} = \Delta_{\underline{x}}^{s}  [R_{2s}] \Big|_{\underline{x}=0},
\]
where the second equality follows from the fact the $\Delta_{\underline{x}}^{s}  [R_{2s}]$ is a constant polynomial. This establishes formula (\ref{Ann}).


\bibliographystyle{abbrv}

\end{document}